# Overcoming the limitations of NMR Field Probes: A Novel Integrated Sensor Utilizing Pre-Polarization for (Ultra) Low Field MRI

Pavel Povolni[1*°], Dominique Goerner[1,2°], Praveen Iyyappan Valsala[1,3], Lukas Gebert[2], Georgiy Solomakha[1], Nicolas Kempf[1], Felix Glang[1,4], Judith Samlow[1,5], Ruben Schnitzler[2], Ingmar Kallfass[2], Kai Buckenmaier[1*], Klaus Scheffler[1,3]

[1] High-Field Magnetic Resonance Center, Max Planck Institute for Biological Cybernetics, 72076 Tübingen, Germany
[2] Institute of Robust Power Semiconductor Systems ILH, University of Stuttgart, 70569 Stuttgart, Germany
[3] Department for Biomedical Magnetic Resonance, University of Tübingen, 72076 Tübingen, Germany
[4] Institute of Biomedical Imaging, Graz University of Technology, 8010 Graz, Austria
[5] Hospital Diospi Suyana, 03100 Curahuasi, Apurímac, Peru

° these authors contributed equally
* Corresponding authors: pavel.povolni@tuebingen.mpg.de, kai.buckenmaier@tuebingen.mpg.de

Abstract

Access to magnetic resonance imaging (MRI) remains severely limited in low- and middle-income countries, especially in *sub-Saharan Africa*, despite rising rates of non-communicable diseases. Low-field MRI presents an affordable, locally developable diagnostic solution, but its performance is constrained by magnetic field instability.

We present a novel NMR field probe designed to overcome these challenges using a rapid non-adiabatic switch-off of a pre-polarization field resulting in precessing spin magnetization. Achieved by first use of high-voltage silicon carbide transistors operating in controlled avalanche breakdown, it measures the Larmor frequency without prior field knowledge—unlike conventional probes. This capability is crucial during magnet development with often unknown fields, allowing early detection of magnet issues, and offering an urgently needed tool for magnet design and image-quality improvement. Validated from 1 mT to 45 mT—up to 1,000 times stronger than similar systems—its low-cost, modular design supports replication, upgrades, and enhanced field control, helping expand global MRI access.

# 1. Introduction

Almost 10 years ago, the United Nations (UN) proclaimed its 17 Sustainable Development Goals (SDGs) targeting the sustainable development of the global community at the economic, social, and environmental levels by 2030[1,2].

However, recent studies indicate that these goals are significantly off track given current progress and that a much greater willingness to act, especially in industrialized nations, is needed[3]. However, given the limited resources available, it is necessary for scientists to identify and address "tipping points" that enable significant system-wide changes, with the *Nature Research* group in particular recognized this need[3,4]. Scientists tasked by the UN are continuously monitoring the progress of the SDGs and has proposed prioritizing especially the goals of global health (SDG3) and education (SDG4), since healthy and well-educated people, for example, increase local productivity and can thus positive impact other SDGs[4]. There is a clear link between these two SDGs in the context of the development and use of medical devices. According to a report by the World Health Organization (WHO), there is a severe shortage of medical devices, especially in low- and middle-income countries (LMICs)[5]. The African continent, and in particular the sub-Saharan region, which currently consist primarily of LMICs, plays a key role in this context, since financial resources of the healthcare system for the purchase and operation of medical devices are very limited.

The issue of restricted access to medical devices is especially apparent in one of the most complex diagnostic imaging technology: magnetic resonance imaging (MRI), which is utilized primarily for the diagnosis of non-communicable diseases such as cancer, cardiovascular diseases, and dementia. Due to the positive development of rising life expectancy in sub-Saharan Africa, the number of people suffering from these diseases is increasing sharply[6–9]. As another example, child mortality due to hydrocephalus, a disease detectable by MRI, could be significantly reduced[6] – directly enforced by SDG3.2, which aims to reduce the current under-5 mortality by ~3 times to below 25 deaths/1000 births[1,2].

However, the significant challenge lies in the enormously high cost of advanced high-field MRI (HFMRI) with a main magnetic field $B_0 \geq 1.5T$ which led to a severely limited accessibility in sub-Saharan countries[10–15]. For instance, Ogbole et. al. reports that West Africa with a population of over 370 million, has access to only 84 MRI (HF and LF) scanners (approx. 0.22 scanner per million inhabitants), while Japan, the leader in this field, has a 230 times higher prevalence[15]. This finding is consistent with the results of various studies, including more recent publications[10–15] and shows a significant problem directly contradicting SDG3 (health). Due to the high complexity and requirements of HFMRI equipment, it is not feasible to install already used devices (from HICs) in sub-Saharan Africa[6].

In contrast to HFMRI scanners, specialized low-field MRI (LFMRI) scanners use magnets with a strength around 50mT for only a small fraction of the costs. These are significantly smaller in size and have low requirements for local infrastructure, no need for cryogenics and may also be developed locally[16–20]. Before we discuss the special features and challenges for the (medical) use of these devices, we would like to briefly touch on the training of local scientists: A significant challenge is the so-called "*brain drain*", which refers to the loss of well-educated individuals to HICs through emigration, resulting in substantial shortages, particularly in the healthcare system (e.g. 20,000 doctors and nurses leave Africa annually)[21]. By improving local infrastructure and job market, consistently implementing the SDGs (e.g., SDG4.4 "Increase the number of young adults with technical skills for employment/jobs/entrepreneurship"[1]) well-trained people can be employed locally[14,22,23]. The open source software and hardware approach is a key way of promoting this, as for example shown by Tsanni et al. in a great example from zoological research in Ghana[24]. If the open science approach is consistently implemented by potentially better equipped research laboratories in HICs, it will directly enable further development to meet specific local requirements in LMICs and is, example given, also endorsed by the UN in its "*Roadmap for Digital Cooperation*"[25].

Due to the significantly lower hardware requirements of LFMRI devices, an open hardware approach was consistently pursued in the development of these devices, enabling them to be successfully assembled and tested in workshops in Uganda, for example[17]. Nowadays permanent magnet-based MRI have proven to be particularly advantageous[17,26–28]. Prior to the dominance of these LFMRI scanners, electromagnets utilizing ohmic conductors were used[16,29]. There is a trade-off between these two systems, since electromagnets can be designed to produce a very homogeneous and drift-stable magnetic field at lower static magnetic fields $B_0$ and permanent magnets achieve a significantly higher Signal-Noise-Ratio (SNR) due to higher $B_0$ but suffer from inhomogeneous and drifting magnetic fields, which represent the most significant technical challenge impacting image quality[19]. At present, permanent magnet-based systems are preferred due to their higher SNR[18]. However, emerging hyperpolarization methods are showing initial good in-vivo results with electromagnets with polarization levels close to those of HFMRI devices and could lead to greater future use of these simpler magnets[30–33].

A fundamental problem for the entire LFMRI research is the lack of highly precise magnetic field sensors: Permanent magnet-based systems require extensive shimming post-construction and electromagnets require a precise alignment of the individual coils for achieving optimal magnetic field homogeneity. To ensure optimal image fidelity, recording of actual imperfect gradients, or potential dynamic $B_0$-drift during a scan is required. For HFMRI, special NMR-based tools are available: commercial sensors are used during construction[34], and self-built[35,36] or commercial systems[37,38] are used as field camera during scans. However, access to commercial systems is limited due to enormous high costs (may be more than the cost of a LFMRI alone).

Developing NMR-based sensors for LFMRI is challenging because of the small point-like measurement volume and the intrinsically low signal $U$, which scales quadratically with the field strength ($U \sim B_0^2$). For instance, compared to 3T-HFMRI, only 1/3,600 at 50mT[26], or 1/210,000 at 6.5mT[29] of the signal remains. Additionally, a large dynamic range is required for field monitoring of the LFMRI setup, especially when the exact field strength is not known in advance. Consequently, currently only completed LFMRI systems can be evaluated with high precision[39] and not during construction (where errors may be fixable).

In order to overcome these limitations of LFMRI, we have developed a novel LFMRI field probe based on an excitation method called non-adiabatic switch-off[40]. An integrated strong pre-polarization field perpendicular to $B_0$ increases the spin polarization significantly above the thermal polarization at $B_0$. Switching it off abruptly, prevents the adiabatic following of the magnetization as the field decays. As a result, the magnetization begins to precess around the $B_0$-axis of the magnet, which is measurable using a receiving coil (Rx coil).

The biggest advantage of this technique is that it operates without any prior knowledge of the magnetic field, unlike conventional excitation schemes which require the exact Larmor frequency to achieve on-resonant condition. Instead, simply switching off the magnetic field quickly enough is sufficient to detect the Larmor frequency.

In order to facilitate this non-adiabatic switch-off, a new operating point of high-voltage n-type metal oxide field effect transistors made of silicon carbide (SiC-nFET) was utilized for the first time. This approach involved deliberately provoking an avalanche breakdown of the SiC crystal allowing very high voltages above 2000V to be safely provided in a compact design, causing the pre-polarization field to collapse with very strong ramps of $\Delta B = -57.6 T/ms$.

Compared with previous implementations of non-adiabatic switch-off used in geophysics at the Earth's magnetic field[41,42], our novel concept achieves 1000-fold faster ramp-down times enabling field measurements up to 3 orders of magnitude higher. Also, this is the first fully integrated sensor of its kind, to the author's knowledge.

Furthermore, the hardware can also be used with a conventional RF excitation that includes standard adiabatic pre-polarization, and has therefore been named *hybrid*.

As part of this work, the functionality of the field probe is demonstrated in an Ultra LFMRI (ULFMRI) device (electromagnet, $B_0 = 1, 3, 5 mT$), as well as a LF Halbach permanent magnet ($B_0 = 45 mT$).

In the aforementioned study, the WHO has highlighted that patents in the field of medical device technology severely restrict access to these devices in LMICs[5]. While there may be potential commercial applications, our understanding is that unrestricted access to the necessary sensing technology is also part of this. Consequently, great emphasis was placed on avoiding the use of special tools as far as possible and making the entire setup truly accessible within the framework of "open hardware"[23].

## 2. Results

### i. Non-Adiabatic Switch Off the Pre-Polarization Field

The core of the sensor is the excitation method using non-adiabatic switching off a strong pre-polarization field. The pre-polarization field $B_p^x$ (along the x-axis) is oriented perpendicular to the static magnetic field $B_0^z$ (along the z-axis) and to the Faraday Rx coil $B_{Rx}^y$ (along the y-axis, see Figure 1a). In contrast to an adiabatic pre-polarization pulse, where the spin magnetization is realigned along $B_0$ after the completed pulse (with higher polarization level), a rapid non-adiabatic switch-off leaves a residual transverse magnetization $M_\perp$ if the adiabatic condition is violated (see Figure 1b)[40]:

$$\gamma |\vec{B}(t)| \gg \left|\frac{d}{dt} \angle (\vec{B}(t), \vec{B_0})\right| \; | \; \forall t \quad (2.1)$$

with $\vec{B}(t) = \vec{B_p}(t) + \vec{B_0}$ and the $\gamma$ gyromagnetic ratio. Note that the operator $\angle$ denotes the angle between $\vec{B}$ and $\vec{B_0}$. The rotating remaining $M_\perp$ inductively generates a measurable voltage $U$ in the Rx coil. The resulting $M_\perp$ after switch-off was numerically simulated using the Bloch equation[43]. In the ideal case of $B_p^x \gg B_0^z$ and a negligible switch-off time $t_p \to 0$, $M_\perp$ is proportional to $B_p^x$, and since a Faraday detection coil is used $U$ becomes $U \sim B_p^x B_0^z$, but is significantly reduced for slower switch-off times (see the detailed analysis in the Supplementary Information (SI) Chapters 1&2).

Povolni, Goerner, et. al. Overcoming the limitations of NMR Field Probes

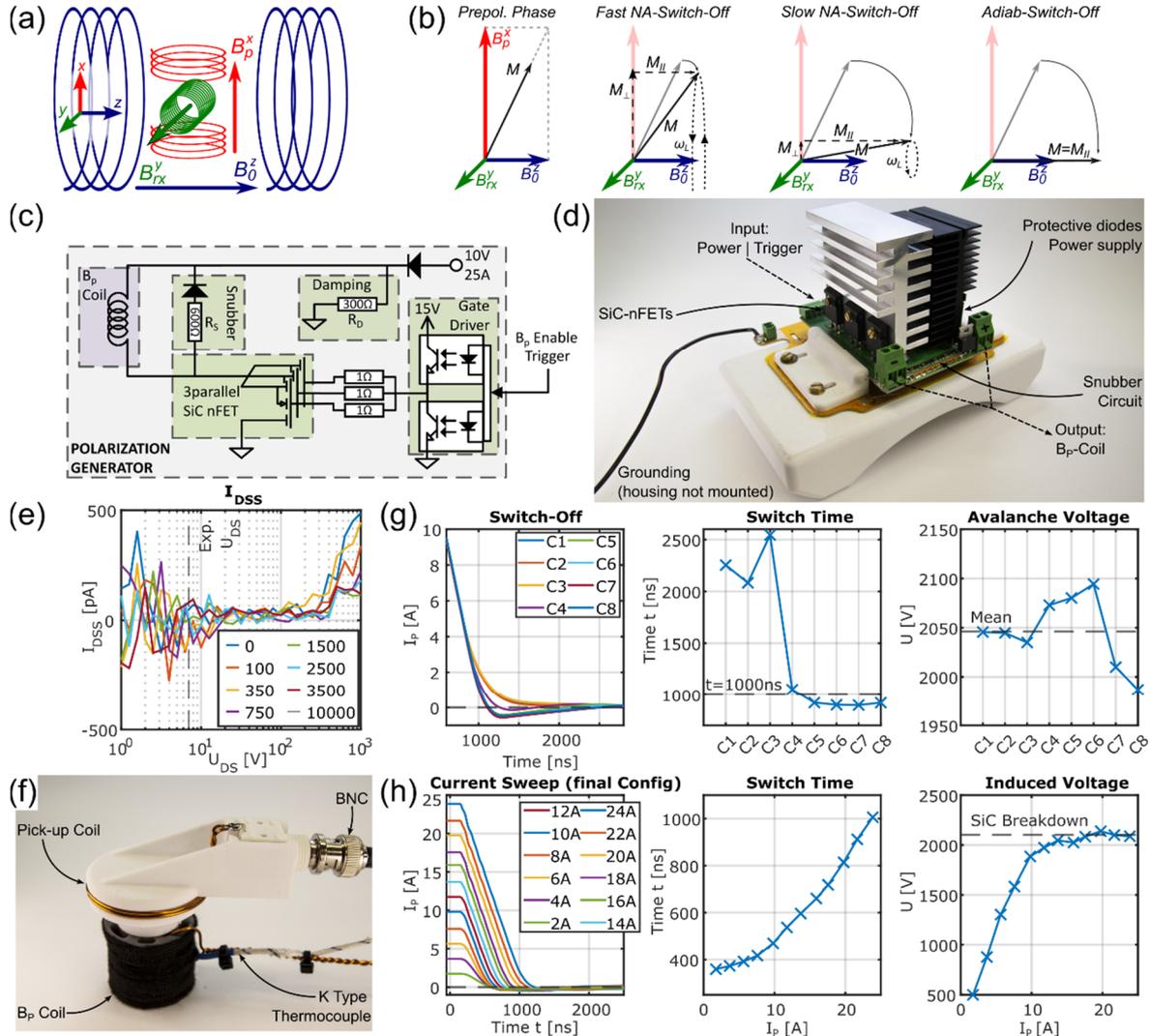

**Figure 1:** Overview of the non-adiabatic switchoff and the used hardware. **(a)** The three magnetic fields $B_p$ (along the x-axis) of the pre-polarization coil, $B_{rx}$ (along the y-axis) of the receiving coil, and $B_0$ (along the z-axis) of the examined magnet are oriented right angled. It is essential to ensure that $B_p$ and $B_{rx}$ are geometrically decoupled to protect the Rx electronics. **(b)** Schematic representation of the transition from the pre-polarization phase to the excited phase (in the case of non-adiabatic (NA) switchoff) or to the initial orientation along $B_0^z$ (in the case of adiabatic switchoff with increased polarization). **(c)** Block diagram illustrating the pre-polarization circuit, showcasing the three parallel SiC-nFETs, which guarantee full operational mode even if a single SiC-nFET fail (also known as *n+1* principle). **(d)** Photograph of the assembled pre-polarization generator mounted on a 3D-printed base. The grounded top housing is not shown. **(e)** Measured reverse current $I_{DSS}$ at applied drain-source-voltage $U_{DS}$ after the aging test consisting of 0/100/…/10,000 measurement cycles. **(f)** Photograph of the designed low-cost pick-up coil for characterizing the shutdown process. **(g)** Investigations of the shutdown process for different configurations of the pre-polarization circuit (set via resistor pair $(R_S|R_D)$ in Ω): C1 (300|225); C2 (300|300); C3 (300|open); C4 (400|300); C5 (500|225); C6 (500|300); C7 (600|225); C8 (600|300)). **(h)** Characterization of the final setup for different pre-polarization currents resulting in a dissipated power ramp of $\Delta P < 55 kW/\mu s$.

Based on simulations the switch-off duration should be $t_p \leq 1\mu s$ for maximizing the available signal in typical (U)LFMRI devices.

In order to switch-off $B_p^x$, a novel pre-polarization generator circuit (pre-polarization current $I_p \leq 25A$, coil inductance $L_p \approx 80\mu H$) was designed to dissipate the energy of the polarization coil as quickly as possible (see Figure 1c-d). The relationship $dI_p/dt = -U_p/L_p$ applies to the current ramp, with the coil voltage $U_p$. This implies that very high voltages in the kilovolt range (referred to as high voltage HV) are required for the aforementioned coil inductance and current $I_p$ (defining $B_p^x$). The HV for this sensor is provided

5 

by the intentionally triggered avalanche breakdown of HV SiC-nFETs (maximum voltage rating $U_{DSmax} = 1.7kV$). Following the abrupt pre-polarization current switch-off, the coil's self-induction causes the voltage across the SiC-nFET to rise until it transitions into a sudden avalanche breakdown ($U_{BD} \approx 2.1kV$) providing a linear discharge of the polarization coil.

It is a standard practice to test SiC-nFETs for their (repetitive) avalanche capability[44] which helps to determine the shutdown behavior in the case of a device failure[45] and to simplify circuits by avoiding extra protection components[46]. To author's knowledge, this sensor represents the first application in which a SiC-nFET is specifically operated in avalanche mode as the desired operating point, especially in this power and voltage class.

Avalanche-induced crystal damage inside the SiC-nFET, which leads to an increased reverse current $I_{DSS}$, is particularly critical because even when switched off, $B_p^x > 0$ applies, which distorts the sensor's precision. $I_{DSS}$ was examined in an aging test with up to 10,000 avalanches, where no damage was detected (shown in Figure 1e, more details in SI Chapter 4).

The non-adiabatic switch-off can be precisely tuned using a snubber circuit to ensure rapid switching without oscillation or setting it to adiabatic pre-polarization[42]. The electrical measurement of this highly dynamic shutdown ($dI_p/dt = -26kA/ms, du/dt = 2.1kV/150ns$) typically involves the use of galvanically decoupled differential probes in combination with high-performance oscilloscopes, which are only available in specialized laboratories. In order to ensure accessibility even in laboratories with limited resources, an easy to replicate solution has been used using an inductively coupled secondary coil in combination with a conventional oscilloscope (see Figure 1f). As illustrated in Figure 1g, the measured switch-off processes for $I_p = 19A$ are shown for different circuit configurations resulting in a needed trade-off between the shutdown duration and the current undershoot. Bloch simulations involving the short undershoot showed no measurable influence on $M_\perp$, so the configuration with the fastest possible switch-off was chosen (resulting in $t_p = 896ns$). The avalanche voltage is $U_{BD} > 2kV$, which matches the expected voltage[47]. The switching behavior of this configuration is demonstrated at varying currents $1.7A \leq I_p \leq 23.9A$ in Figure 1h, resulting in switch-off times from $359ns \leq t_p \leq 1006ns$. For $I_p < 12A$, the stored energy is not sufficient for a SiC-breakdown and will be dissipated in the snubber circuit only.

In this configuration, for ULFMRI with $B_0^z \leq 5mT$, the adiabatic condition (2.1) can be significantly violated for all tested $B_p^x \leq 70mT$, resulting in measurable $M_\perp$. For LFMRI with $B_0^z \approx 50mT$, the adiabatic condition is generally satisfied. In the numerical Bloch simulation, however, a very weak $M_\perp$ remains visible and is measurable if the sensor concept is designed accordingly. The comparatively high Larmor frequency of approximately $f_0 \approx 2MHz$ is advantageous in this case, as the induced voltage in the Rx coil is proportional to $f_0$.

ii. Design of the sample, polarization coil, and receive coil for maximum sensitivity

The NMR signal strength is predictably low at (U)LFMRI. To ensure that the sensor remains functional for the targeted wide range of $\Delta B_0 = 1 - 50mT$, the Rx coil and pre-polarization coil were optimized using simulation. First, the sample volume was set, with a trade-off made between maximum volume (higher NMR signal due to more polarized spins) and the strongest possible pre-polarization field (limited to $L \leq 80\mu H$ and $I_p \leq 25A$). For $B_0 \leq 5mT$, $V = 1.2ml$, and at $B_0 = 45mT$, $V = 1.5ml$ was selected due to a smaller expected signal. Pure water was used as sample achieving highest possible spin density (signal strength). The polarization coil has been optimized to achieve the strongest possible field at high homogeneity under the given boundaries resulting in a multilayer solenoid coil (see Figure 2a and b). For an RF-excitation with adiabatic pre-polarization (*hybrid mode*), the innermost layer was used as Tx $B_1$-coil and the remaining layers for pre-polarization.

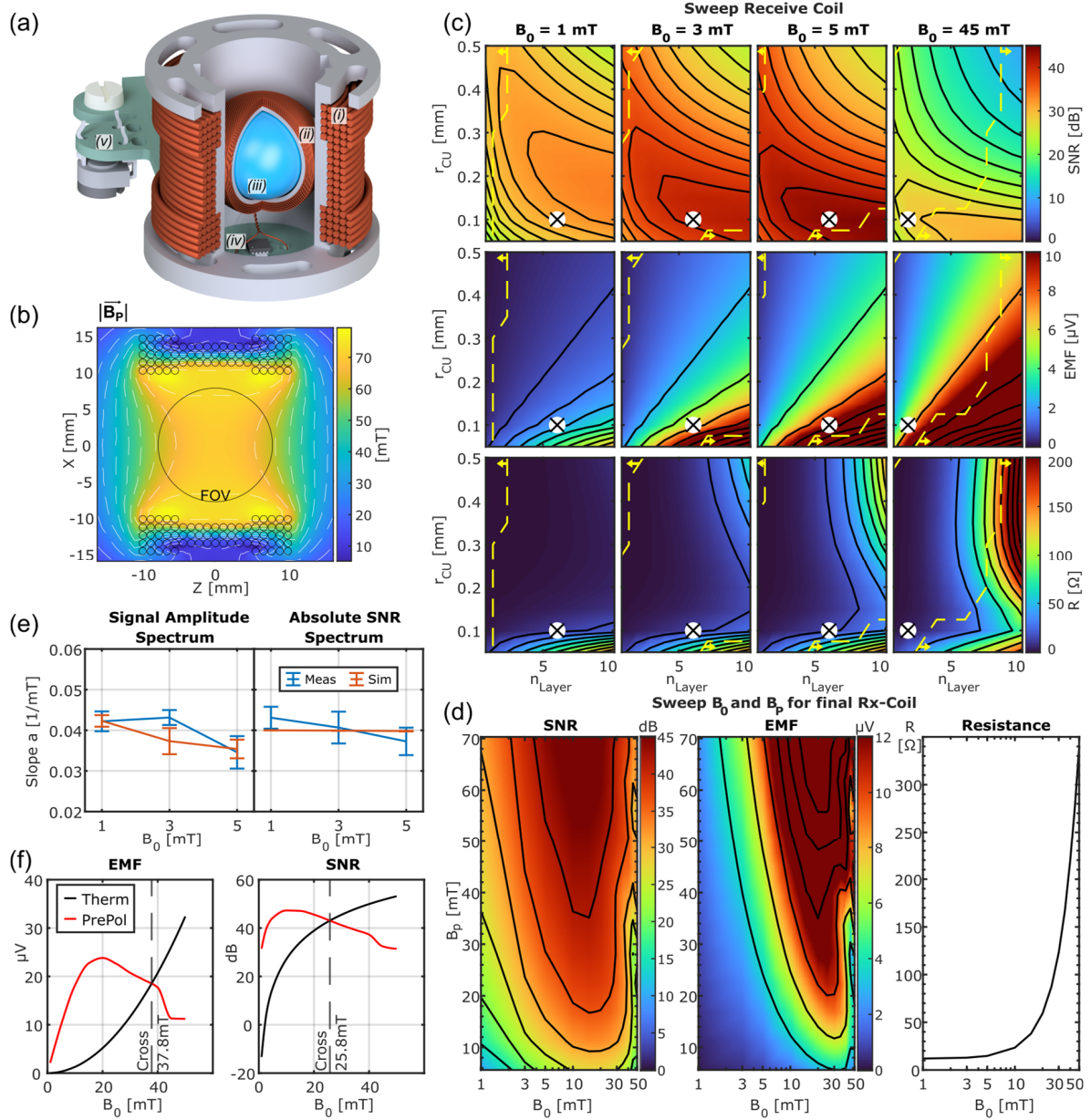

**Figure 2:** Design of the pre-polarization and Rx coil. **(a)** CAD rendering of the assembled frontend with *(i)* the outer pre-polarization coil, *(ii)* the inner Rx coil, *(iii)* the water sample located in an impregnated sample container, *(iv)* a closely positioned 3D Hall sensor, and *(v)* the frontend printed-circuit-board (PCB). **(b)** Resulting $B_p$-field from FEM simulation of the optimized pre-polarization coil ($I_p$= 20A). **(c)** Optimization of the Rx coil for a simulated NMR signal at $B_0$=1/3/5/45mT (which were also experimentally measured in this work). The copper wire radius was varied in a range 0.1mm≤$r_{Cu}$≤0.5mm. The number of winding layers was varied in a range 1≤$n_{Layer}$≤10. In conjunction with geometrical constraints both parameters define the total number of axial turns. The calculated *SNR* (top), the calculated *EMF* (middle), and the coil-resistance *R* (bottom) are shown. The yellow boundary (line with arrows) indicates the range in which the Rx coil **cannot** be feasibly matched to Z=200Ω (optimum input impedance for LNA). For $B_0$≤5mT the pre-polarization $B_p$=61.9mT ($t_p$=918ns) and for $B_0$=45mT the pre-polarization $B_p$=70.3mT ($t_p$=1074ns) was used. The white marker indicates the chosen Rx coil geometry. **(d)** Expected *SNR*, *EMF*, and *R* for the final design at different $B_0$ and $B_p$ combinations, where the measured $t_p$ were used for each set $B_p$. For easier comparison, the ULF configuration of the Rx coil is used for all fields. Note the logarithmic scaling of the x-axis. **(e)** A qualitative comparison between simulated and measured linear fit of *SNR/EMF*($B_p$) with $B_0$ held constant. $a(B_0)$ corresponds to the linear slope in the formulas $EMF_{norm}(B_p)=EMF(B_p)/EMF_{max}=a(B_0)B_p+OS_{EMF}$, resp. $SNR_{norm}(B_p)=SNR(B_p)/SNR_{max}=a(B_0)B_p+OS_{SNR}$. $OS_{EMF}/OS_{SNR}$ indicates the intercept. The NMR spectra were normalized to the maximum value due to slight variations in Rx coil geometries between simulation and experiment. **(f)** Comparison of the simulated *EMF* and *SNR* between the excitation methods using non-adiabatic switchoff ($B_p$=70mT) and conventional RF excitation with thermal polarization (same optimized Rx coil used). The observed anomaly at $B_0$≈40mT in the switch-off method potentially results from inaccuracies in the measured $T_2^*$ values, which are affected by $B_0$ inhomogeneities at higher fields. For thermal polarization a constant $T_2^*$ was chosen for simplicity.

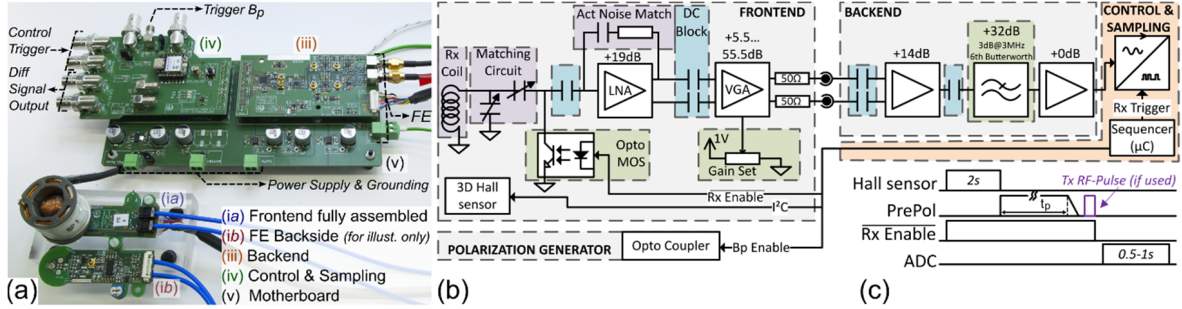

**Figure 3:** Overview of the receive electronics. **(a)** Photography of the entire Rx hardware as a modular architecture consisting of four PCBs. Note, that the flipped frontend *(ib)* is shown only for illustration purposes and not used in the experiment. **(b)** Block diagram of the Rx electronics. **(c)** Measurement sequence used during the shown experiments. The Hall sensor could operate significantly faster than 2s. Here it was switched on, configured, read out and switched off during its time block with enough delay time to have it switched off before the strong $B_p$ is switched on. Rx Enable is active-low signal, therefore logical high switches the photocoupler (OptoMOS) on and grounds the LNA RF input.

The shape of the Rx coil was optimized by solving the Bloch equations numerically and calculating the induced signal for the various coil configurations (see Methods for details).

The calculated *SNR*, electromotive force *EMF* (voltage signal amplitude), and resistance *R* of the coil for varied geometries are shown in Figure 2c. The expected *EMF* increases significantly across all magnetic fields with more turns (many layers with small wire radius), which, however, results in higher resistances due to the total copper length limiting the achievable *SNR*. The highly frequency-dependent resistance (due to skin effects) also reduce the *SNR* at $B_0 = 45mT$ significantly below the value at $B_0 = 1mT$. Based on these results, optimized ULF ($B_0 \leq 5mT$) and LF ($B_0 = 45mT$) configurations of the Rx coil were build (see markers in Figure 2c).

Using these configurations and the measured pre-polarization ramps, the expected *EMF*, *SNR* and *R* were calculated for $1mT \leq B_0 \leq 45mT$ and $5mT \leq B_p \leq 70mT$ (see Figure 2d). As with the initial Rx coil optimization, it is evident that pre-polarization is *SNR*-wise especially beneficial at low $B_0$. Due to impedance boundary constraints, the sensor may use the ULF configuration up to $B_0 \leq 10mT$ and the LF configuration for higher fields. The linear increase of *SNR* and *EMF* over $B_p$ was compared for $B_0 \leq 5mT$ with the normed measured results (see Figure 6a for measurement details and SI Chapter 5 for calculations). The simulated slope aligns closely with the measured slope, as demonstrated in Figure 2e, highlighting the accurate calculation of noise sources in the simulation for *SNR* estimation.

A simulated comparison of the *SNR* of this non-adiabatic excitation and a conventional RF excitation with thermal polarization reveals that non-adiabatic excitation is clearly advantageous up to a limiting magnetic field of $B_0 = 26mT$ ($B_0 = 38mT$ when considering *EMF*) which is shown in Figure 2f. Above this field the conventional RF excitation generates higher *SNR*

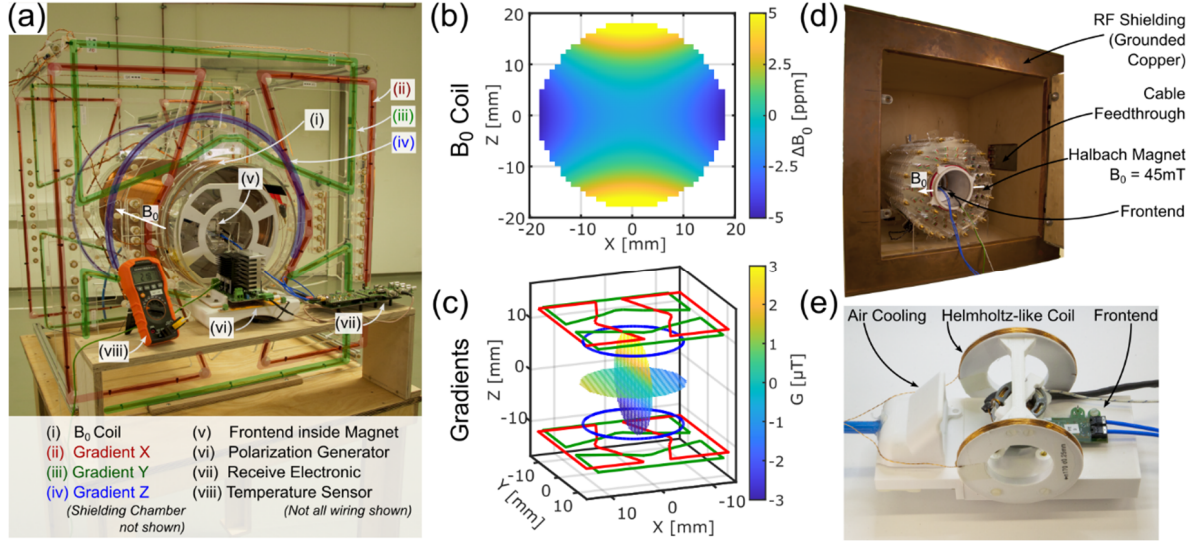

**Figure 4:** Overview of the magnetic fields investigated. **(a)** The used ULFMRI based on a homogeneity-optimized electromagnet for $B_0$ and two planar X/Y gradients and a Maxwell coil as the Z gradient. An exemplary experimental setup is shown with the frontend positioned on a slider inside the magnet and the backend and pre-polarization generator placed outside the magnet. **(b)** Simulated $B_0$ homogeneity $\Delta B_0$ in parts-per-million (ppm) of the ULFMRI magnet shown in (a). **(c)** Simulation of gradient efficiency at a gradient current of $I_G=1A$. Only the magnetic fields of the X and Z gradient are shown (Y is analog to X, but rotated by 90°). The gradient coordinates are scaled by a factor of 1/25 for this plot. Note, that the plot is rotated by 90° around the x-axis compared to (a) for illustrational purposes. **(d)** Photograph of the Halbach magnet inside the grounded RF shielding chamber. The frontend is positioned inside the magnet. Other electronics is not shown, but is positioned inside the RF shielding chamber during the experiments. The shielding may be designed in another way with e.g. conductive cloths or copper meshes, since the frequencies in MHz-range can be shielded easily. **(e)** Photograph of the slider (usually positioned inside the Halbach magnet), accompanied by the Helmholtz-like offset coil for $B_0$ field variation. The frontend is positioned in the isocenter of the offset coil and the slider is positioned, so that the sensor sample is approximately at the magnet isocenter.

— but only when $B_0$ is known accurately for an on-resonant excitation.

### iii. Design of the Receiver Electronics

A modular architecture was chosen to streamline debugging and upgrading if needed (shown in Figure 3). The frontend, with mounted polarization coil and Rx coil (see Figure 2a) followed by a low noise amplifier (LNA), is positioned within the magnetic field. An additional Hall sensor in the center of the coil pair is used for rough Larmor frequency estimation to simplify spectrum analysis, as this estimated frequency determines the range for digital bandpass filtering[48]. The backend with an active anti-aliasing filter remains outside the magnetic field. In the shown experiments, the NMR signal is digitized using available commercial hardware like oscilloscopes. A fully integrated solution is also provided for replication with the design files in the shared repository, but requires extensive knowledge of embedded coding for functionality.

### iv. Measurement environments and examined magnets

The examined magnets are shown in Figure 4. The ULFMRI based on a solenoid and gradient system was used for $B_0 = 1/3/5\ mT$[49]. For $B_0 = 45mT$, an unshimmed Halbach permanent magnet with an inhomogeneity of $\Delta B \approx 370 ppm$ (inside the sensor sample) and an integrated $B_0$ offset coil was used (see Figure 4e), which, however, significantly decreases $B_0$ homogeneity[48]. The individual used frontends are listed in Table 1.

**Table 1:** Overview of the used setups. The Rx coils were wound manually, resulting in a random winding pattern and slightly different coupling inductances between the individual conductor loops. This leads to different inductances *L* and quality factors *Q* despite the same number of turns. There is a slight deviation when impedance matching, due to the discretized capacities on the frontend PCB and the strong inductive coupling with the polarization coil. For the pre-polarization coil the inductivity *L*, the DC Resistance $R_{DC}$ and the calculated pre-polarization efficiency $k_{Bp}$ used in the relation $B_p=k_{Bp}I_p$ is shown. For the conventional RF excitation the most inner layer of the pre-polarization coil was used.

| Setup | | 1mT | 3mT | 5mT | 45mT | 1mT RF Tx |
|---|---|---|---|---|---|---|
| Rx Coil | *V* (Water) | 1.2ml | 1.2ml | 1.2ml | 1.5ml | 1.2ml |
| | L | 811µH | 950µH | 771µH | 64µH | 811µH |
| | Q | 17.5 | 18.2 | 18.5 | 13.2 | 17.5 |
| Polarization coil | L | 80µH | 80µH | 81µH | 80µH | 49µH |
| | $R_{DC}$ | 153mΩ | 150mΩ | 156mΩ | 153mΩ | 122mΩ |
| | $k_{Bp}$ | 2.812mT/A | 2.812mT/A | 2.812mT/A | 2.812mT/A | 1.832mT/A |
| RF Transmit | $B_1$ Coil L | x | x | x | x | 6.9µH |
| | $B_1$ Coil $R_{DC}$ | | | | | 46mΩ |
| | $B_1$ Coil $k_{B1}$ | | | | | 0.684mT/A |
| Rx Electronics | Gain (Overall) | 110 dB$_V$ (except Figure 5a) | 110 dB$_V$ | 105 dB$_V$ | 102.5 dB$_V$ | 107.5 dB$_V$ |
| | Sampling | 500ms 16bit | 500ms 16bit | 500ms 16bit | 1000ms 12bit | 500ms 16bit |

### v. NMR Measurements in Ultra Low Field for $B_0 \leq 5mT$

Figure 5 shows exemplary NMR spectra from the ULF magnet setup with the aim of demonstrating the functionality of the field probe over a wide range away from its theoretical ideal working point (related to the matching frequency). Close to the matching frequency (vertical dashed line in *SNR* plots of Figure 5) the maximum *SNR* is expected. From this ideal frequency, the sensor operates over a comparatively wide range of $B_0$, which can be especially extracted from the *SNR* plots in Figure 5. According to this evaluation, the measured working range $\Delta f_{OP}$ is $\Delta f_{OP} \in [-15|+50]\%$ for $B_0 = 1mT$, $\Delta f_{OP} \in [-22|+15]\%$ for $B_0 = 3mT$ and $\Delta f_{OP} \in [-25|+17]\%$ for $B_0 = 5mT$ from the ideal matching frequency.

The FWHM at $B_0 = 3/5mT$ is well below $\Delta f < 10Hz$ (resp. $\Delta f < 30Hz$ for $B_0 = 1mT$) resulting in an achievable sensitivity of $f_{sens} < 26ppm$ (resp. $f_{sens} < 235ppm$ for $B_0 = 1mT$, see SI Chapter 9 for details), demonstrating the effectiveness of the build pre-polarization setup with minimizing transient disturbances[50].

The dependence of the pre-polarization phase on the non-adiabatic switch-off was investigated by varying the field strength from $0mT \leq B_p \leq 67.5mT$ (constant pre-polarization time $t_p = 6s$) (results shown in Figure 6a) and in another experiment by varying the pre-polarization time $1s \leq t_p \leq 10s$ (constant pre-polarization field $B_p \approx 61mT$) (results shown in Figure 6b). In principle, ohmic losses in the polarization coil resulting in heating the water sample needs to be reduced by minimizing $B_p$ or duration $t_p$, with $B_p$ having a stronger influence (energy loss $E_{loss} \sim B_p^2 t_p$). As $B_p$ increases, the measured amplitude increases linearly (as expected since $M_\perp \sim B_p$ for $B_p \gg B_0$). Already small pre-polarization fields $B_p \geq 10mT$ are sufficient to measure an NMR spectrum. As $t_p$ increases, the signal amplitude follows an exponential behavior with the time constant being the NMR relaxation time $T_1$. The measured $T_1$ times are within the expected range and can be used for future and more precise simulation[32]. As seen, already short pre-polarization durations of $t_p > 2s$ are sufficient to measure the Larmor frequency. Although the electronic performance of the pre-polarization generator varies significantly in both sweeps due to different working points, the determined FWHM is constant across the sweep range, highlighting the high spectral resolution of the sensor, even with very efficient pre-polarization phase. In Figure 6b the measured amplitudes for $B_0 = 3mT$ dominates, which is in contrast to the simulation and the $B_p$ sweep experiments. Due to the pre-polarization generator's design, there is an unknown delay between the actual excitation and the sampling start (more details in Methods and SI Chapter 7), which can be the reason for the observed differences in amplitudes. However, NMR detection functionality remains operational.

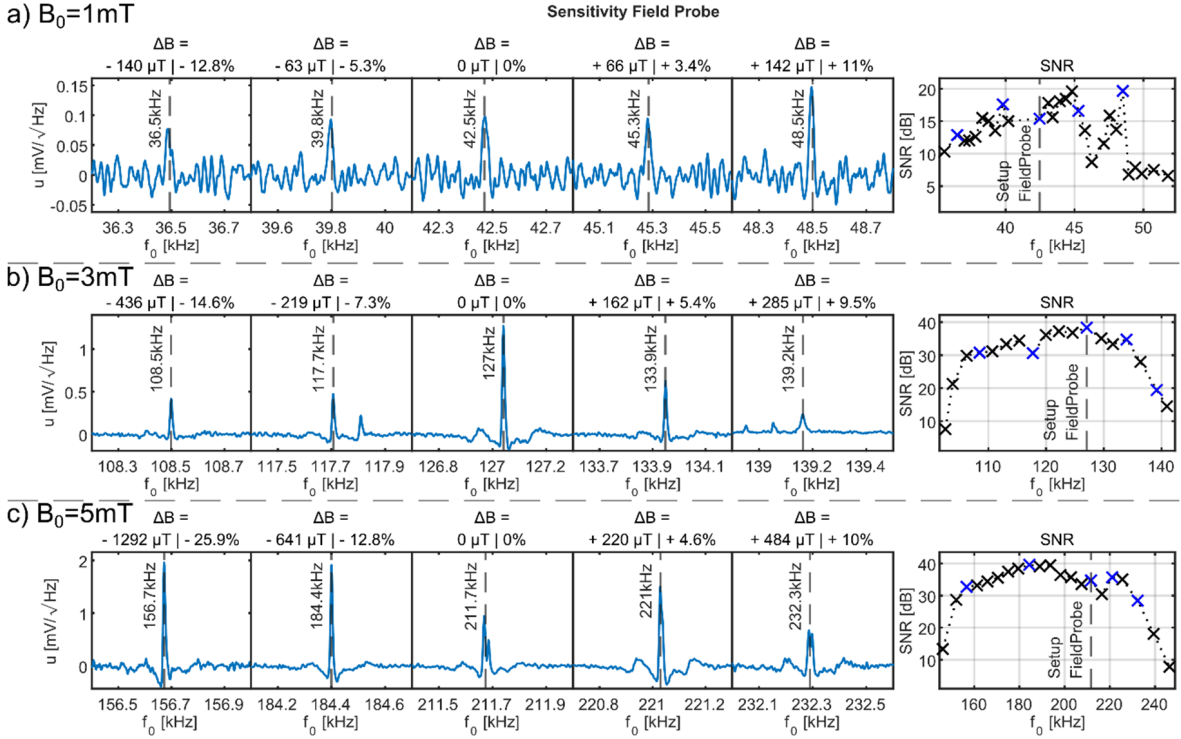

**Figure 5:** Exemplary phase corrected NMR spectra in the ULF for determining the sensing ranges of the assembled sensors. For every $B_0$ setup three NMR spectra were measured and averaged. The voltage related *SNR* is derived from the NMR spectra and shown on the right side each, where the blue colored markers indicate the exemplary spectra shown on the left side. **(a)** Measured NMR spectra at $B_0$=1mT ($B_p$=62mT, $t_p$=6s). Note, that the set gain of the Rx electronics was significantly reduced to *G*=105dB due to strong noise sources resulting in railing of the Rx electronics. This led to smaller signal amplitudes and therefore smaller *SNR* (e.g. due to increased quantization noise of the sampling). Despite the noisy environment, the Larmor frequency may be detected without any issues. **(b)** Measured NMR spectra at $B_0$=3mT ($B_p$=61mT, $t_p$=6s). The observable side peaks are not related to the NMR signal and are probably influenced by noise sources in the laboratory. **(c)** Measured NMR spectra at $B_0$=5mT ($B_p$=60mT, $t_p$=6s).

The implemented design offers the possibility of being easily adjusted for high $B_0$ resolution (strong $B_p$, long $t_p$, long $T_R$), for high temporal resolution (strong $B_p$, short $t_p$, short $T_R$), or for energy efficient long-term drift measurements

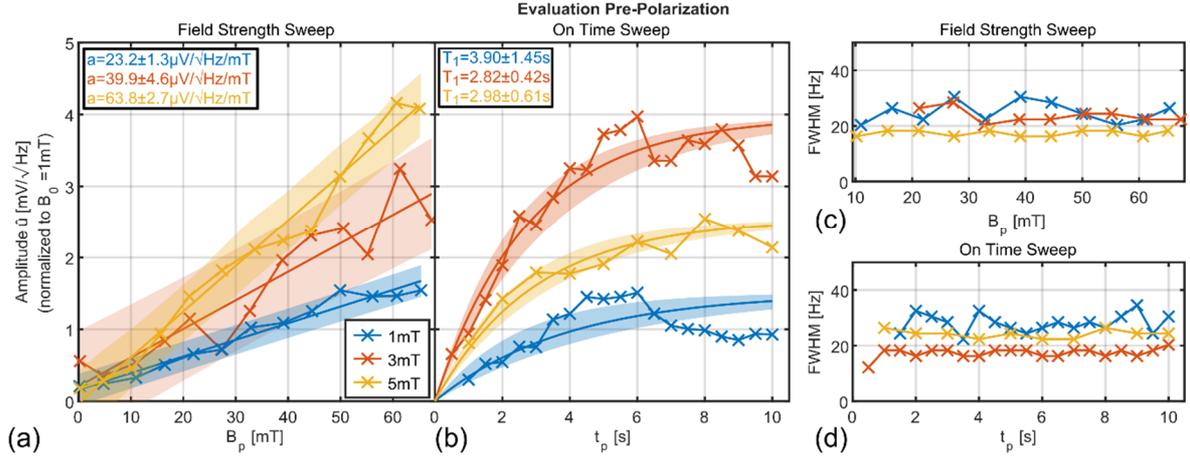

**Figure 6:** Investigation of the influence of the pre-polarization field strength $B_p$ and duration $t_p$ on the measured signal amplitude. To ensure comparability, all FID data were normalized to the total gain G=110dB set at $B_0$=1mT. For every configuration, one NMR spectrum was acquired (no averages). **(a)** Resulting signal amplitude for a swept pre-polarization field strength in the range 0mT<$B_p$<70mT ($t_p$=6s). The signal amplitude is linearly fitted with $\hat{u}=aB_p+OS_u$ resulting in the shown slopes $a$. **(b)** Resulting signal amplitude for a swept pre-polarization duration in the range 0.5s<$t_p$<10s ($B_p$=61mT). The signal amplitude is exponentially fitted with $\hat{u}(t)=\hat{u}_{max}(1-\exp(-t/T_1))$ resulting in the shown relaxation times $T_1$. **(c)** Calculated FWHM for the acquired NMR spectra during the pre-polarization field strength sweep in (a). **(d)** Calculated FWHM for the acquired NMR spectra during the pre-polarization duration sweep in (b).

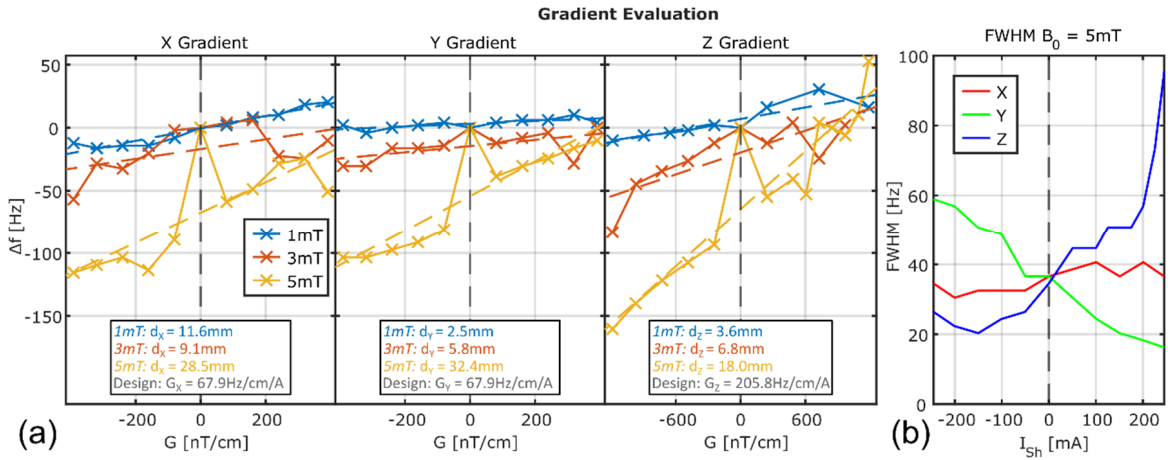

**Figure 7:** Investigation of the performance of the ULFMRI gradients. **(a)** Measured frequency drift with an applied X/Y/Z gradient in ULF (in all cases $B_p$=61mT, $t_p$=6s), with slight differences in the positioning accuracy of the frontend. Only one measurement was performed (no averages, so simulating a fast acquiring mode for dynamic mapping). The X/Y gradient was set in a range $\Delta G_{X,Y}$=±414nT/cm (11 equal steps) and the Z gradient in a range $\Delta G_Z$=±1256nT/cm (11 equal steps). The calculated offset positions $d_{X,Y,Z}$ are shown in the textboxes. **(b)** Measured FWHM for experiments at $B_0$=5mT for swept gradient currents. Only one gradient was switched on at a time. One spectrum was acquired (no averaging). The strongest influence on the linewidth (indicating inhomogeneity inside the sample) is caused by the Y and Z gradients (which indicates stronger field inhomogeneity in these directions).

(weak $B_p$, long $t_p$, long $T_R$).

The designed sensor is capable of detecting even the weak field variations of the installed gradients. For evaluation, the gradients were switched on with a direct current (DC) and the magnetic field was measured. The sensor was located close to the isocenter with an unknown positioning tolerance between the individual field setups. The resulting frequency drifts are illustrated in Figure 7a. Here, only one position was evaluated, which, however, is enough to determine the relative sensor position to the center of gradients (see SI Chapter 6 for details). This indicates a substantial deviation, likely attributable to improper alignment of gradients. In the measurement process, first the gradients were switched off, and only then the gradients were switched on and swept. With already experienced temporal field drift this may explain the unexpected maxima at switched off gradients at $B_0 = 5mT$ (see SI Chapter 8 for further field drift evaluation). The first order shimming possibility of the gradients, resp. the measured verification of it, is shown exemplary for $B_0 = 5mT$ in Figure 7b, where the FWHM-change is shown for varying gradients.

### vi. Hybrid measurement with RF excitation in Ultra Low Field

As illustrated in Figure 8a, the hybrid excitation, where adiabatic pre-polarization is followed by a conventional on-resonant RF excitation, is demonstrated for $B_0 = 1mT$. Testing at this low-field strength is considered a critical challenge to the sensor concept, since matching of the Rx and Tx coils is difficult at low frequencies using large capacitances and small induced NMR signals.

The effect of the pre-polarization was evaluated by sweeping the pre-polarization field, resp. using only the thermal polarization. The successful impact of pre-polarization is evident in the increasing signal amplitude

### vii. NMR Measurements in Low Field at $B_0$ = 45mT

In Figure 9, the measured NMR spectra of measurements inside an unshimmed Halbach magnet at $B_0 \approx 45mT$ are shown. Three measurements were taken for each operating point ranging from $\Delta B_0 = -1228\mu T$ to $\Delta B_0 = 0\mu T$. The expected $f_0$ value plotted is estimated from the known offset coil constant. For each operating point, there is at least one functioning measurement with a mean relaxation time of $T_2^* \approx 0.12ms$ (averaged $FWHM = 4028Hz$) in the investigated volume, which fits the estimated magnet inhomogeneity from a previous Hall sensor based mapping extended by the inhomogeneity introduced by the offset coil[48]. The results of the consecutive measurements at the same operating point demonstrate a substantial drift of the magnet (shown in Figure 9b). On average, the repetition time is $\overline{T_R} = 29.2s$ (manual triggering), resulting in an averaged frequency drift of $\Delta f = 835Hz$. Due to the significant frequency drift, averaging the spectra is not recommended as it broadens the measured peak and consequently reduce *SNR*. Moreover, averaging is not necessary, as *SNR* is sufficient in a single measurement already.

Due to the very short $T_2^*$, the resolution of the $B_0$ mapping is limited in this unshimmed case, but is still sufficient for an initial shimming and significantly more precise than Hall sensors (here $\Delta_{NMR} \approx 0.1\%$ compared to the build in Hall sensor $\Delta_{Hall} \approx 1\%$, see SI Chapter 9 for details). Significantly longer $T_2^*$ are expected for shimmed magnets. Since the measured NMR peak frequency corresponds exactly to the set offset field, it can be assumed that the influence of the switch-off mode is negligible to the sensor accuracy.

One potential explanation for some of the missing spectra could be attributed to the switching of the pre-polarization generator itself. Due to the short relaxation time $T_2^*$, it is essential that the sampling start promptly after finishing the non-adiabatic excitation. As the avalanche does not always initiate at the same time, even for an adapted sequence with a minimum possible delay, the signal might have already decayed too much until the sampling starts (see SI Chapter 7 for a detailed analysis).

## 3. Discussion

In this work, we have developed a functional, cost-effective, low-field probe capable of rapidly and autonomously examining the magnetic field of an unknown (U)LFMRI-magnet in the range $1 \leq B_0 \leq 50mT$. The central component is the newly designed pre-polarization generator, which enables rapid non-adiabatic switch-off of the pre-polarization field, leaving a transverse magnetization $M_\perp$ without any additional on-resonant excitation. The key novelty here was the first use of a high-voltage SiC-nFET at a previously unused dedicated operating point, where avalanche breakdown was purposefully triggered at $U_{BD} \approx 2.1 kV$, resulting in very strong current ramps of $\Delta i = -26.4 kA/ms$ ($\Delta B = -57.6 T/ms$).

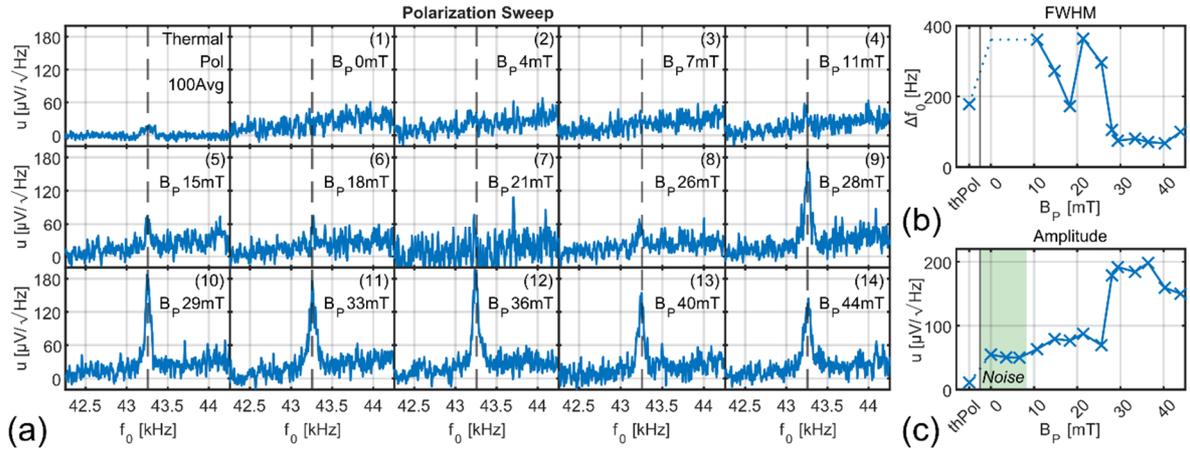

**Figure 8:** Hybrid excitation using a combination of pre-polarization with adiabatic switch-off and oscillating RF excitation at $B_0$=1mT. **(a)** Exemplary NMR spectra at different $B_p$ (absolute value of the complex spectrum is shown). Two measurements are shown without pre-polarization (and therefore thermal polarization only): For the first experiment the pre-polarization generator was switched completely off and 100 measurements were averaged. The second experiment was part of the $B_p$-sweep experiments starting with the pre-polarization coil current $I_p$=0A. For the $B_p$-sweep experiments four measurements were averaged (therefore higher noise level compared to first experiment). **(b)** The FWHM derived from (a). **(c)** The amplitude $u$ derived from (a). Note, that the first 3 measured amplitudes of the $B_p$-Sweep correspond to the measured noise level.

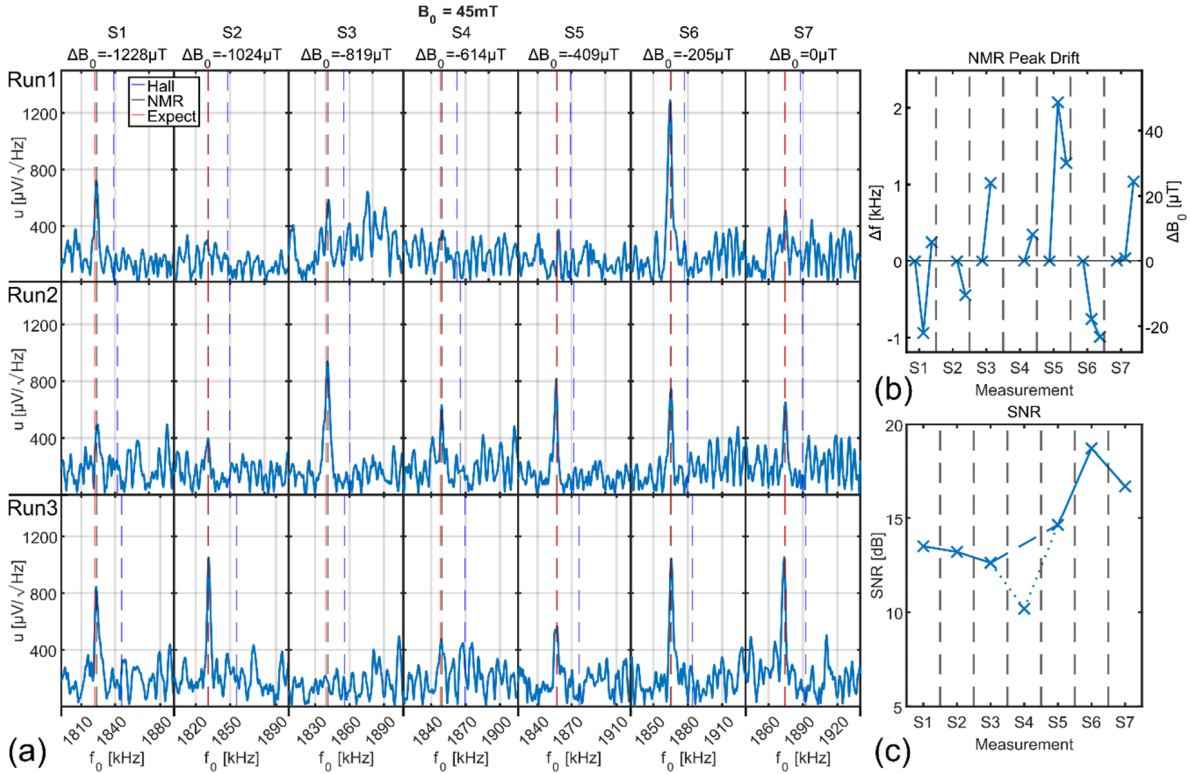

**Figure 9:** Measured NMR spectra for an unshimmed Halbach magnet at $B_0$=45mT. The field probe was tuned to $f$=1.852MHz and pre-polarization set to $B_p$=70.3mT for $t_p$=6s. **(a)** Exemplary selection of acquired NMR spectra (absolute value of the complex spectra is shown) in a varying offset field of -1.228mT<$\Delta B_0$<0mT in seven equal distant steps. The three vertical measurements are consecutively measured taken immediately after each other. Data was not averaged. '*Hall*' shows the measured value of the integrated Hall sensor, '*NMR*' shows the averaged Larmor frequency of the three measurements, and '*Expect*' shows the expected Larmor frequency due to the set of the additional coil. The measurements were taken at approximately 30-second intervals. **(b)** Frequency drift $\Delta f$ derived from (a) at the same operating points (missing points are non-detectable NMR peaks). **(c)** SNR of the most successful measurement (strongest amplitude) derived from (a).

Non-adiabatic switch-off has been proven to be an elegant solution for use in magnetic field sensors and was already successful in earth magnetic field based NMR[41,42], or close to zero-field for physical experiments[51]. However, in this work we were able to expand the measurable field strength by 3 orders of magnitude using the novel pre-polarization generator. Pre-polarization is advantageous in ultra-low-fields because it significantly enhances spin polarization, leading to higher signal level and therefore resulting in reduced requirements for the receive electronics. However, with permanent magnets-based MRI with typical $B_0 > 40 mT$, this gain is no longer substantial. At these field strengths another major advantage is still the frequency independent excitation, so no a priori knowledge of the measured field is needed. Conventional NMR-based field probes can only measure the exact Larmor frequency via a Tx pulse frequency sweep[34–36] since the Tx pulse must be resonant with the Larmor frequency. This is a time-consuming technique due to small signals and huge matching network adjustments. This requirement is eliminated by the presented non-adiabatic switch-off concept.

In principle, any other high-voltage source could be used for the pre-polarization circuit to enable the needed current ramps. However, due to the same earth potential and to cover possible insulation faults, extensive personal protection measures (e.g., insulation monitors, interlocks etc.) must be implemented by trained high-voltage technicians ($U > 1kV$ is medium-voltage-level and life-threatening). Additionally, appropriate high-voltage sources are costly and considered specialized equipment, typically unavailable in laboratories of LMICs.

The implemented solution provides an elegant and minimalistic option to achieve the needed high-voltage at minimal financial cost. As the high-voltage is generated locally within the transistor housing, a simple touch protection of the PCB and the coil is sufficient. The stored energy inside the coil $W_P < 25 mJ$ is not dangerous since it is similar to a typical electrostatic discharge of the human body, which one, for example, typically experiences during winter months[52].

With the implemented fail-safe design, which was achieved through parallelization, the selection of appropriate SiC-nFETs, the substantial cooling system, and the aging tests, it is safe to assume that more than 10,000 measurement cycles per transistor are easily possible. Possible damage is detectable in further developments by implementing a reverse current measurement or could be remedied by replacing the (cheap) transistors after certain cycles. Even faster switch-offs would be possible by integrating novel developed 3.3kV-SiC-nFETs in the build PCB, which enables even higher $B_0$ to be measured.

By design, an accurate matching of the Rx coil is not required, as it is only necessary for noise matching with the first LNA (not for noise free signal amplification by tuning). Currently, the frontends are not exchangeable and the frontend must be rematched or replaced for different field strengths. To minimize expenses, the frontend was designed to use as few components as possible so multiple ones can be built. Additionally, the RF path was designed to allow rematching even when fully assembled.

Future developments could automate the matching adjustments with electronically adjustable capacitors, enabling a single frontend to cover a broader field range.

For HFMRI, NMR field probes are used for improving image quality by measuring the actual time-varying magnetic field (influenced by imperfect gradients, eddy currents etc.) during the MRI examination[53]. The field probe presented here extends this function to (U)LFMRI: The drift of Halbach magnets, for example, is a known problem that was also recorded by our experiments (see Figure 9). In our unshimmed magnet, the drift was exceptionally high, which likely prohibits any MR-imaging. With this field probe, however, it can be solved by active shimming using the sensor output as a controller input (similar setup to[36]). The same is possible for the well-known current drift in actively driven electromagnets[29,31].

Using this sensor, the DC characteristics of a newly designed (and not yet characterized) gradient for an (U)LFMRI can be evaluated without any a priori-knowledge about it, which is especially valuable in the development of non-standard gradient systems, e.g. for specialized (U)LFMR scanners for dedicated human body parts. Using the non-adiabatic excitation, a high temporal resolution is not yet achievable. However, this would be possible with the identical sensor hardware using its hybrid mode in case of a sufficiently stable magnetic field (which should be already the case for MRI imaging).

Water was used as the sample to maximize signal. Using the engineered impregnation of the sample container a dry-out was successfully prevented. Additionally, the coils were mechanically designed so that the sample can be removed from the polarization coil and refilled. Presently, the pre-polarization time causes long repetition times and limits the measurement duty cycle. To achieve faster and more efficient measurements, a future development should initially use a *CuSO₄*-doped water sample and, in a subsequent step, a solid (rubber) sample to significantly shorten $T_1$[35]. It is also feasible to use fluorinated samples, as fluorine has a comparable gyromagnetic ratio but remains unaffected by proton MRI-imaging. The volume could also be reduced due to the achieved *SNR*, although it is currently of the same order of magnitude as a voxel in LFMRI[27].

In summary, we have designed a functional field probe that could be particularly helpful to the emerging (U)LFMRI community.

Particular emphasis during design was placed on ensuring that this hardware is as easy to replicate and upgradable as possible – even with constrained resources[5,25], resulting in a modular electronic architecture and the avoidance of any special tools, parts or expertise. For the actual sampling, existing commercial hardware was used in the presented measurements to significantly reduce the complexity of the system. Due to the low frequency, a simple and cheap oscilloscope is sufficient and may be more accessible than a fully integrated solution.

All hardware and software, along with detailed instructions (including the ULF magnet for testing), is available as open hardware, enabling direct improvements to all LFMRI scanners already built and installed, as well as the development of new hardware and sequences. The straightforward design enables local scientists and technicians in LMICs to advance MRI hardware and sequences for addressing specific local challenges.

The development of such specific and optimized LFMRI technologies will enable the use of patient-oriented and location-independent low-field scanners to meet unmet clinical needs in various healthcare locations worldwide and has the potential to further democratize MRI for low- and middle-income countries.

# 4. Methods

### i. Simulation and Mechanical Design of the Pre-Polarization Coil

The pre-polarization field strength and possible switch-off times result in a tradeoff of the winding pattern since high number of windings create a strong field but also high inductance. With the design limitations of the pre-polarization generator and especially with the limited and not adjustable avalanche voltage, the inductance was chosen to be of higher importance. One important additional geometrical constraint is the possibility to replace the inner Rx coil, since preliminary experiments showed fast drying of the sample due to insufficient tightness.

Different variations of the polarization coil have been calculated using FEM (implemented in the open source software FEMM[54]) and the best option was chosen. The final polarization coil is a multi-layer solenoid coil with an efficiency of $k_{Bp} = 2.812 mT/A$, $L = 80 \mu H$, using an enameled copper wire with a radius of $r_{Cu} = 0.5mm$. In the axial direction, the coil has 19 windings and 3 layers, with the outer 5 windings on the inside and outside each having one more winding (Figure 2a). By following a prescribed winding specification, it is possible to wind the coil uniformly and avoid gaps between the layers. The connecting wires are approximately 30cm long twisted wire. The coil former is 3D printed using Stereolithography (SLA), and due to the thermosetting plastic, it is particularly temperature-stable. Another printing technology that uses a suitable temperature-resistant material can also be used. To ensure the stability of the windings during the winding process, they were affixed with superglue.

The pre-polarization coil was modified for the hybrid mode with adiabatic pre-polarization and RF Tx pulses. The bottom copper layer is separated as a *B₁*-coil with efficiency $k_{TX} = 0.684 mT/A$ and matched to $Z \approx 16\Omega$ at $f = 43.26 kHz$ using a single-sided matching network. Due to the low frequency, ideal 50Ω matching is not necessary. The remaining outer layers collectively constitute the pre-polarization coil with efficiency $k_{BP} = 1.832 mT/A$. For both sub-coils enameled copper wire with $r_{Cu} = 0.5mm$ was chosen.

### ii. Simulation and Mechanical Design of the Receive Coil

In order to optimize the receive coil design, the Bloch equation was solved numerically in a MATLAB script to calculate the transversal magnetization of the sample[43]. A steady state magnetization before the switch-off was assumed (quasi infinitely long pre-polarization). Using the well-proven reciprocity principle, the induced voltage can be calculated, if the *B₁*-characteristics of the Rx coil is known for a unit current prior[55]. The *B₁*-characteristics of the Rx coil was calculated using Biot-Savart's Law[56]. This process was recalculated for all configurations of *B₀*, sample volume *V*, *Bₚ*, switching time *tₚ*, and all Rx coil parameters (number of layers $n_{Layer}$, number of windings *n* and wire radius $r_{Cu}$).

The resistance was calculated using the DC resistance of overall copper length given the number of windings and radius and by considering the frequency dependent skin[57] and proximity effects[58]. This overall resistance was used to calculate the Johnson-Nyquist noise (using the real bandwidth of the Rx electronics). Further noise sources of the Rx electronics and environmental noise were neglected for simplicity (see SI Chapter 3 for details). The achievable *SNR* was estimated using this noise level and the calculated induced voltage signal and is therefore overestimated significantly (since noise will be higher in reality).

An additional boundary condition involves the possibility to match the Rx coil to an impedance of $Z = 200\Omega$ (optimal input impedance with minimum noise figure of the used LNA). Based on the simulation results, the following parameter combinations were selected for the Rx coils: For $B_0 \leq 5mT$ with $r_{Cu} = 0.1mm$ and $n_{Layer} = 6$ (ULF configuration) and for $B_0 = 45mT$ $r_{Cu} = 0.1mm$ and $n_{Layer} = 2$ (LF configuration).

The Rx coil was wound onto a spherical sample container so that the windings are positioned geometrically as close as possible to the signal source. For the ULF configuration, the optimum target is 570 total turns. However, due to an irregular pattern resulting from the manual winding process and the spherical coil former, only 360 counted turns could be wound. According to the simulation (see Figure 1c) this should only have a minor influence on achievable *SNR*. The windings were stabilized with superglue. A balanced matching circuit consisting of parallel fixed and adjustable, nonmagnetic capacitors was adjusted as close as possible to $Z = 200\Omega$ using a VNA at the desired Larmor frequency.

### iii. Sample Container Design

The spherical sample container was 3D printed using a FDM printer from PETG with a nozzle diameter of $d_N = 0.1mm$ and was double-sealed with a polymer impregnation (DIAMANT Polymer GmbH, Moenchengladbach, Germany) in an immersion bath. The wall thickness is $d_w = 0.6mm$. The internal volume is a sphere with a volume of $V = 1.2ml$ (ULF configuration) or $V = 1.5ml$ (LF configuration). The container is filled with water as a sample using a syringe. Afterwards, the hole was sealed by melting it using a soldering iron. The impregnation was tested for three days in a vacuum chamber at a pressure $p = 250mbar$ (water boiling point $T = 70°C$, room temperature $T = 25°C$). This experiment shows no measurable water loss.

### iv. Receive Electronics

According to the Friis-formula, the noise characteristics of the first amplifier are crucial. The single-ended LNA AD8331 (Analog Devices, Norwood, USA), typically used in ultrasound imaging, was used with a gain of $G = 19dB$ and $NF = 1dB$ ($Z_{in} = 200\Omega$). A subsequent variable gain amplifier (VGA, same package) allows manual fine adjustment of the overall Rx gain by feeding an adjustable voltage (manually set by voltage divider with potentiometer, see Figure 3). The VGA output is differential enabling a more noise-resistant signal transport outside the magnet. An active sixth-order Butterworth anti-aliasing filter with a cut-off frequency of $f_{3dB} = 3MHz$ is placed on the backend PCB, which is positioned outside the magnetic field. The filter is designed single-ended for easier manufacturing (fewer tolerance-affected components). A differential-to-single-ended converter is positioned in front of the filter. The voltage-related total amplification of the input voltage signal ranges from $G = 72dB$ (gain 3980) to $G = 120dB$ (gain 1,000,000), depending on the VGA setting. The VGA gain is optimized to prevent noise-induced railing in subsequent amplifiers. A photocoupler (TLP3431, Toshiba Electronic, Tokyo, Japan) short-circuits the Rx coil during the pre-polarization phase. The circuit on the frontend was evaluated using the software Advanced Design System (Keysight Technologies, Santa Rosa, USA) to ensure circuit stability in noisy environments. The filter was simulated using the software LTSpice (Analog Devices). The output of the VGA was stabilized with a series resistor $R = 50\Omega$, and the output of the filter with $R = 1k\Omega$ against capacitive loads (like coaxial wires). For the measurements in this work, $B_0 \leq 5mT$ was sampled with a PXI-6363 card ($t_s = 0.5s$, $Q = 16bit$, $f_s = 2MSps$, National Instruments, Austin, USA) and $B_0 = 45mT$ was sampled with a HDO6104-MS oscilloscope ($t_s = 1s$, $Q = 12bit$, $f_s = 10MSps$, Teledyne LeCroy, Thousand Oaks, USA). Any other oscilloscope with a resolution of $Q \geq 12bit$, a sampling rate $f_S \geq 5MSps$ and a memory $m \geq 2.5MS$ is sufficient. The signal is then evaluated in post-processing on a PC using a MATLAB (The MathWorks Inc, Natick, USA) script. An alternative option is available in the design files, which includes an integrated analog-to-digital-converter (ADC, LTC2225, Analog Devices, $Q = 12bit$, $f_s = 10MSps$) and evaluation with a FPGA (iCE40UP5K, Lattice Semiconductor, Portland, USA). Please note that this option requires some experience in embedded coding. An additional 3D Hall sensor ALS31300 (Allegro Microsystems, Manchester, USA) with a measuring range of $B_{X/Y/Z} = \pm 50mT$ ($\pm 0.7\%$) has been installed in the center of the pre-polarization coil and is used in particular for rough Larmor frequency estimation to simplify spectrum analysis. The Hall sensor is only powered during its readout and switched off otherwise. The sequence control and readout of the Hall sensor is managed by a microcontroller (ESP32C3, Espressif Systems, Shanghai, China). The power supply for all sub-boards is consistently noise-optimized, with digital and analog components having their own galvanically separate battery-powered power supply. The individual potential is only brought to the same potential at a single neutral point (star grounding of all PCBs).

### v. Pre-Polarization Electronics

The pre-polarization generator is based on the SiC-nFET G3R160MT17D (GeneSiC Semiconductor, Torrance, USA) with a maximum voltage of $U_{DSmax} = 1.7kV$ (breakdown voltage is approx. $U_{BD} \approx 1.3 U_{DSmax}$)[47] an on-resistance $R_{DSon} = 160m\Omega$ and can handle a non-repetitive avalanche energy of $E_{AS} = 158mJ$. Three identical SiC-nFETs from the same batch are connected in parallel (cost ~10€ per piece), thereby reducing ohmic losses during the pre-polarization phase and ensuring reliability of the sensor. The SiC-nFETs are operated with the gate driver ACPL-P343 ($I_{pk} = 4A$, $U_G = 0V/+15V$, Broadcom Inc., San José, USA). The pre-polarization duration can be freely adjusted and is limited solely by the ohmic losses of the polarization coil (temperature measurement with type K sensor). In our measurements, the coil was cooled using active air cooling with compressed air. The semiconductors are passively cooled with heat sinks and exhibit negligible heating. A laboratory power supply with $U \leq 20V$ and $I \leq 45A$ (PE1643, Philips, Amsterdam, Netherlands) is used as a high-current source. In order to trigger the avalanche breakdown, no free-wheeling diodes were used for protection against the voltage peaks due to self-induction of the pre-polarization coil[45,59]. The SiC-nFET was subjected to 10,000 cycles in an additional aging test, and its characteristic curve was examined using a Curve Tracer B1505 (Keysight Technologies) with a high-current and high-voltage extension (further details in SI Chapter 4).

In the developed circuit, the shutdown process can be precisely adjusted to the load inductance via a snubber circuit consisting of a series resistor $R_S$ and a pn-diode ($U_{BD} = 2.2kV$, DNA30E2200PA, IXYS Corporation, Milpitas, USA)[42]. A damping resistor ($R_D$) prevents oscillations by dissipating remaining coil energy after the stopped avalanche (see Figure 1c). This network enables to adapt the circuit for an adiabatic switch-off ($R_S = 0\Omega$) and a non-adiabatic switch-off ($R_S > 300\Omega$). For high $R_S$, the current undershoots during the switch-off. To achieve fastest possible switchoff, $R_S = 600\Omega$ and $R_D = 300\Omega$ were selected. For in-situ characterization, an inductively coupled pickup coil was placed in close proximity to the polarization coil (distance between the centers of the coils $l \approx 27.6mm$). The pickup coil has two turns of enameled copper wire ($r_{Cu} = 0.5mm$), a coil diameter of $d_{PU} = 38mm$, negligible DC resistance, and is terminated in series with $R = 50\Omega$. An older oscilloscope ($BW = 70MHz$, $f_s = 2GSps$) without any special galvanic protection measured the voltage of the pickup coil and therefore the current ramp. The coupling inductance to the primary coil has been measured to $L_{mut} \approx 450nH$. The largely linear behaving current ramp indicates a constant coil voltage, which can be calculated directly by the relationship $U_{BD} = L|\Delta i/\Delta t|$ using the measured linear slope.

The duration of the avalanche-powered switch-off process was determined to be largely constant with a deviation within the measurement tolerance. However, the delay between the trigger and the completed shutdown process is inherently not consistent due to the circuit, the driver and slight deviations in the avalanche start. For an optimized sequence for measurement at $B_0 = 45mT$, this results in an arbitrary measured delay between $2\,\mu s < t_d < 26\mu s$ (see SI Chapter 7 for details).

vi. Hybrid Excitation (Classical NMR Approach)

For a conventional NMR experiment with RF excitation, the polarization coil at the frontend is replaced by the hybrid $B_1$ polarization coil (specified in Table 1), and the pre-polarization generator is set to adiabatic switch-off mode. The maximum pre-polarization field $B_p = 46mT$ (at $I_p = 25A$) switches off in $t \approx 48\mu s$. A waveform generator 33500B (Agilent/Keysight Technologies) with a variable output voltage and pulse length $t = 5ms$ is used as RF source. A set voltage $U_{RMS} = 28.5mV_{RMS}$ corresponds to a flip angle $\alpha = 90°$. The waveform generator is triggered at the corresponding timepoint by the built-in sequencer.

For the hybrid excitation experiments shown in Figure 8, a flip angle of $\alpha \approx 82°$ ($B_1 = 1.07\mu T$ for $t_{Tx} = 5ms$) was set. The pre-polarization field and duration was set to $B_p = 36mT$ and $t_p = 5984ms$ respectively. The frequency of the Tx pulse is set to $f_{Tx} = 43.257kHz$.

vii. Sequences Used

For non-adiabatic shutdown (all $B_0$): 2s Hall sensor measurement, $t_p = 6s$ (generally, if not stated otherwise), 0.1ms pause, 500ms sampling.

For hybrid excitation at $B_0$=1mT: 2s Hall sensor measurement, $t_p = 5984ms$, 10ms pause, 5ms RF excitation, 1ms pause, 500ms sampling.

viii. Signal Filtering and Processing

To keep the sensor's integrated computational power low and thus the hardware inexpensive, complex filtering and fitting of the signal was avoided. Though, in this work the sampled voltage data were processed on a standard computer using MATLAB scripts. Simple algorithms were used to stay compatible with an embedded microcontroller for compatibility. The data was filtered in time domain using a normalized apodization filter by multiplying the measurement data with the following weighting function[60]:

$$a(t) = \frac{1}{\max\limits_{\forall t} a(t)} \sin\left(\pi \frac{t}{T}\right) \exp\left(-k \frac{t}{T}\right) \quad (4.1)$$

with the parameters $k = 15$ and $T = 500ms$ for $B_0 \leq 5mT$, and $T = 2ms$ for $B_0 = 45mT$.

Due to the non-ideal switch-off with varying delay, the first ADC samples during $t \leq 7.5ms$ were not evaluated for $B_0 \leq 5mT$ (all possible interferences have decayed). At $B_0 = 45mT$, this offset is dynamically adjusted in post-processing to follow the actual trigger edge at the gate of the SiC-nFET (also sampled) due to the really short *T₂\** relaxation time (details shown in SI Chapter 7). The resulting truncated data, was digitally filtered using a zero-phase bandpass infinite impulse response (IIR) filter (Butterworth, fourth order, bandwidth $BW = 4kHz$ for $B_0 \leq 5mT$ and $BW = 500kHz$ for $B_0 = 45mT$)[61]. The center frequency was calculated based on to the magnetic field measured by the Hall sensor. The time-domain signal is then fast Fourier transformed to obtain the frequency spectrum.

For the characterization of the presented hardware and the shown NMR spectra in this work, a semi-automated phase correction was performed, and the signal was cubic interpolated in the frequency domain to $\Delta f = 0.1 Hz$. The real part of the spectrum is usually evaluated for all measurements, except for the hybrid excitation measurement and the LFMRI Halbach measurements due to low *SNR* (and therefore no phase correction was performed). For stand-alone sensor operation, these extra processing steps such as phase correction and interpolation are unnecessary. Analyzing the absolute spectra is sufficient, as only the peak frequency has to be calculated.

ix. Ultra-Low Field Magnet and MRI System

The central electromagnet is a solenoid coil with diameter $d = 270mm$, length $l = 470mm$ and containing 470 turns of enameled copper wire ($r_{Cu} = 0.5mm$, more details in SI Chapter 10). To optimize homogeneity, a second layer of 30 turns was added to both ends. The coil efficiency at the isocenter is $k = 2.24 mT/A$ with a calculated homogeneity of $\Delta B_0 = 12 ppm$ over a diameter of spherical volume (DSV) of $d = 20mm$ (corresponding $\Delta f = 2.5 Hz$ at $B_0 = 5mT$)[49]. The current source BCS3/12 (12V/3A, HighFinesse GmbH, Tübingen, Germany) was used for $B_0 = 1mT$ and the current source BCS5/75 (75V/5A, HighFinesse GmbH, Tübingen, Germany) was used for $B_0 = 3|5mT$[31]. Previous experiments have shown that these current sources, and thus the resulting *B₀*, drift slightly with the more powerful source performing worse. Using an another ULFMRI system, the current drift of this source was determined to be $\Delta I \approx -0.47 ppm/s$. This means that a drift of $\Delta f = 42.775\ mHz/A/s$ should be expected for this magnet. However, this value may differ considerably depending on the source performance. For a typical sensing time of $T_R = 10s$ in this work, this results in a possible drift of $\Delta f = 0.2 Hz$ (1mT), $\Delta f = 0.6 Hz$ (3mT) and $\Delta f = 1 Hz$ (5mT) per measurement (see SI Chapter 8 for details).

The built-in gradients also serve as first-order shimming. The Z-gradient consists of a 32-turn ($r_{Cu} = 0.2mm$ copper wire) Maxwell coil with a diameter of $d = 462mm$ and a spacing $h = 400mm$, which is identical to the system in[31]. The simulated magnetic efficiency is $G_Z = 483 \mu T/m/A$. The X- and Y-gradients consist of planar coils with 32 turns of $r_{Cu} = 0.4mm$ wire. They measure $A = 612x697mm^2$ and are mounted $h = 595mm$ apart. The design corresponds to the system in[31], but have been scaled up by 20% resulting in a simulated magnetic efficiency of $G_X = G_Y = 159 \mu T/m/A$. The exact gradient accuracy and efficiency is not known. Self-developed, battery-powered current sources with an output of $I_{out} = \pm 250 mA$ were used[31]. The experiment was positioned in a three-layer magnetic shielding chamber (Vacuumschmelze GmbH, Hanau, Germany), with a residual magnetic field $B < 10 nT$ at its center.

x. Halbach Permanent Magnet

The Halbach magnet consists of 536 NdFeB N48 magnets (12x12x12 mm³,

$B_R = 1.395T \pm 25mT$) with a bore size of $d = 160mm$ and a length $l = 300mm$. In a previous experiment, the magnet was mapped using a Hall sensor-based mapping resulting in an inhomogeneity of $\Delta B = 8351 ppm$ in a DSV of $d = 40mm$[48]. Scaled to the sensor's volume, this corresponds to $\Delta B \approx 370 ppm$ ($\Delta f \approx 720 Hz$). The inhomogeneity could be much higher in our setup because of the uncertainty in positioning the sensor.

To vary $B_0$ and prove the sensor performance, an air-cooled Helmholtz-like coil with diameter $d = 72mm$, distance $h = 76mm$ and $n = 170$ turns of copper wire ($r_{Cu} = 0.15mm$) was integrated into the Halbach magnet (illustrated in Figure 4e). The isocenters of both magnets were aligned. The frontend was positioned accordingly, so that the sensor sample was positioned at the isocenter of the combined magnet. Compared to an ideal Helmholtz-coil, the coil distance was doubled here due to geometrical constraints resulting in lower magnetic efficiency. FEM simulation show an efficiency of $k_{BOS} = 2.047\ \mu T/mA$ at the center by adding a significant high inhomogeneity of $\Delta f = 5.349\ Hz/mA$ inside the sensor sample. The overall resistance is $R = 29.4\Omega$. The same current source from the ULFMRI was used (BCS3/12, 12V/3A, HighFinesse). The offset coil was only activated shortly before the NMR measurement (long enough for steady state of the sample magnetization) and was otherwise switched off to reduce the heating of the Halbach permanent magnets. The entire magnet was positioned in a grounded RF shielding chamber. For active cooling purpose, the polarization and offset coil were exposed to an active air flow using compressed air.

# 6. Further Information

### xi. Acknowledgements

P.P. would like to thank S. Eckardt (GreenTeam, University of Stuttgart) for his help in embedded coding. Funding by European Research Council (ERC Advanced Grant No 834940, SpreadMRI) is gratefully acknowledged. K.B. acknowledges funding from the DFG (BU 2694/6-1, BU 2694/9-1, 469366436, 527345502).

### xii. Author contributions

Study conception and design: P.P, D.G., K.B and K.S.
Acquisition of data: P.P, D.G, L.G, R.S and K.B
Analysis and interpretation of data: P.P, D.G, P.I.V, L.G, G.S, N.K, F.G, J.S, R.S, I.K, K.B and K.S
Drafting of manuscript: P.P, D.G and K.B
Critical revision: all authors

### xiii. Data and Code availability

The source code of the software and hardware to reproduce or extend this work are available under
https://github.com/ppolGTMonster/Hybrid_LowField_NMRProbe

# 7. Supplementary Information

## i. Calculation of transverse magnetization after switch-off process

The sample is pre-polarized by a strong polarization field $B_p$, which is then abruptly switched off. Because of the circuit design, the switch-off ramp exhibits a quasi-linear characteristic. In order to proceed with the design of the receive circuit, it is necessary to estimate the remaining transverse magnetization after the successful switch-off. Simultaneously, it is possible to adjust the pre-polarization circuit and coil such that a good trade-off between the $B_p$ field strength, the switch-off duration $t_p$ and the available, measurable magnetization is made. The calculation of the available transverse magnetization is derived in the following sections.

The time varying total spatial magnetic fields are:

$$B_x(t) = \begin{cases} B_p\left(1 - t/t_p\right) & | \ 0 \leq t \leq t_p \\ 0 & | \ t_p \leq t \end{cases} \tag{S1.1}$$

$$B_y = 0 \ \forall t \tag{S1.2}$$

$$B_z = B_0 \ \forall t \tag{S1.3}$$

This results in the differential equation according to Felix Bloch [1]:

$$\begin{pmatrix} \dot{M}_x(t) \\ \dot{M}_y(t) \\ \dot{M}_z(t) \end{pmatrix} = \gamma \begin{pmatrix} 0 & B_z & 0 \\ -B_z & 0 & B_x(t) \\ 0 & -B_x(t) & 0 \end{pmatrix} \begin{pmatrix} M_x(t) \\ M_y(t) \\ M_z(t) \end{pmatrix} \tag{S1.4}$$

Due to the short switch-off time, longitudinal and transverse relaxations are neglected.

This results in a double accelerated coordinate system for the macroscopic spin vector: On the one hand, the effective magnetic field to which the spins are oriented rotates around the y-axis of the laboratory system with time-varying amplitude $B_{eff}$, angle $\varphi$, and angular velocity $\omega_{eff}$:

$$B_{eff}(t) = \sqrt{B_x(t)^2 + B_0^2} \tag{S1.5}$$

$$\varphi(t) = \operatorname{atan}\left(B_x(t)/B_0\right) \tag{S1.6}$$

$$\omega_{eff} = \frac{d\varphi(t)}{dt} = \frac{-B_0 B_p t_p}{B_p^2 t^2 - 2B_p^2 t_p t + (B_0^2 + B_p^2)t_p^2} \tag{S1.7}$$

Additionally, the magnetization precesses around the effective magnetic field $B_{eff}$ (Larmor precession). This precession slows down due to the time-varying magnetic field strength, resulting in the following relationship between precession frequency $\omega_L$ and magnetic field strength:

$$\omega_L(t) = \gamma_{1H} B_{eff}(t) \tag{S1.8}$$

An analytical solution to the differential equation S1.4 cannot be found. One potential approach involves the use of the Magnus expansion, though it should be noted that convergence of the series is not guaranteed (our approximation up to the 5th order was not sufficiently accurate). Therefore, the necessity arises for a numerical solution to the differential equation. The process of solving the differential equation can be time-consuming, particularly when one attempts to sweep $B_0$, $B_p$, and $t_p$ in three dimensions simultaneously. In the present study, a high-performance computer node (512 GB RAM, 2x64-core AMD EPYC 7452 2.35 GHz) was utilized for a duration of several hours to complete the calculation of the magnetization trajectory. However, for the design of the Rx circuit, only the remaining transverse magnetization after the switched off pre-polarization at $t = t_p$ needs to be known. Simplified approximations can therefore be employed to rapidly test a large number of designs, thus facilitating the process of design iteration. Based on numerical solutions, in the range (which should be interesting and practical for (ultra) low field MRI):

$$1 mT \leq B_0 \leq 60 mT$$
$$5 mT \leq B_p \leq 100\ mT$$
$$200\ ns \leq t_p \leq 4800\ ns$$

a quadratic approximation was calculated.

The basis is the following dependence of the transverse magnetization $M_\perp$ remaining at the Rx-start:

$$M_\perp(t_p) = m_{rel}(B_0, B_p, t_p) M_{\perp 0}(B_0, B_p) \tag{S1.9}$$

With transverse magnetization $M_{\perp 0}$ at time $t = 0s$ (steady state) (equation 6.10 [2]):

$$M_{\perp 0}(B_0, B_p) = \rho_0 \frac{\gamma^2 \hbar^2}{4kT} B_{eff}(t=0) \tag{S1.10}$$

with the spin density per unit volume $\rho_0 = 6.68 \cdot 10^{28}$ for water (proton), the gyromagnetic ratio $\gamma = 2.675 \cdot 10^8 s^{-1} T^{-1}$ for water (proton), the reduced Planck constant $\hbar$, the temperature $T$ in Kelvin and the Boltzmann constant $k$.

The following empirical formula ($B_0$ in mT, $B_p$ in mT, $t_p$ in µs) is employed to calculate the magnetization ratio:

$$\begin{aligned} m_{rel}(B_0, B_p, t_p) &= p_1 B_0^2 + p_2 B_p^2 + p_3 t_p^2 \\ &\quad + p_4 B_0 B_p + p_5 B_0 t_p + p_6 B_p t_p \\ &\quad + p_7 B_0 + p_8 B_p + p_9 t_p + p_{10} \end{aligned} \tag{S1.11}$$

where the averaged time period of the Larmor precession is used to separate 3 individual ranges for higher accuracy:

$$t_L = \left( \gamma \sqrt{(0.5 B_p)^2 + B_0^2} \right)^{-1} \tag{S1.12}$$

with the gyromagnetic ratio $\gamma = 42.577\ MHz T^{-1}$ for water (proton).

The calculated parameters are:

|  | Unit | $t_L \leq 0.5 t_p$ | $0.5 < t_L \leq 1.5 t_p$ | else |
|---|---|---|---|---|
| p1 | [1/mT²] | + 4.41e-04 | + 1.27e-04 | + 1.12e-04 |
| p2 | [1/mT²] | - 3.59e-06 | - 1.83e-05 | + 3.97e-06 |
| p3 | [1/µs²] | + 0.0134 | + 0.0141 | - 9.02e-03 |
| p4 | [1/mT²] | + 7.20e-05 | + 5.79e-05 | - 4.28e-06 |
| p5 | [1/µs/mT] | - 9.56e-04 | - 1.38e-02 | + 0.0464 |
| p6 | [1/µs/mT] | - 3.28e-04 | - 2.15e-04 | + 1.43e-03 |
| p7 | [1/mT] | + 0.0397 | - 0.0103 | - 8.57e-03 |
| p8 | [1/mT] | - 5.51e-03 | + 7.01e-03 | + 9.70e-04 |
| p9 | [1/µs] | + 0.137 | - 0.0971 | - 0.0548 |
| p10 | [1] | - 0.990 | + 0.612 | + 1.02 |

The mean discrepancy between the numerical solution and the approximation is 5%, with deviations of up to 30% possible in individual extreme cases (in the case of extremely unusual combinations such as long $t_P$ with weak $B_p$). If needed, this calculated level of accuracy should be adequate for the design of new Rx electronics or Rx coils. The comparison of both calculations is demonstrated in Figure S1.

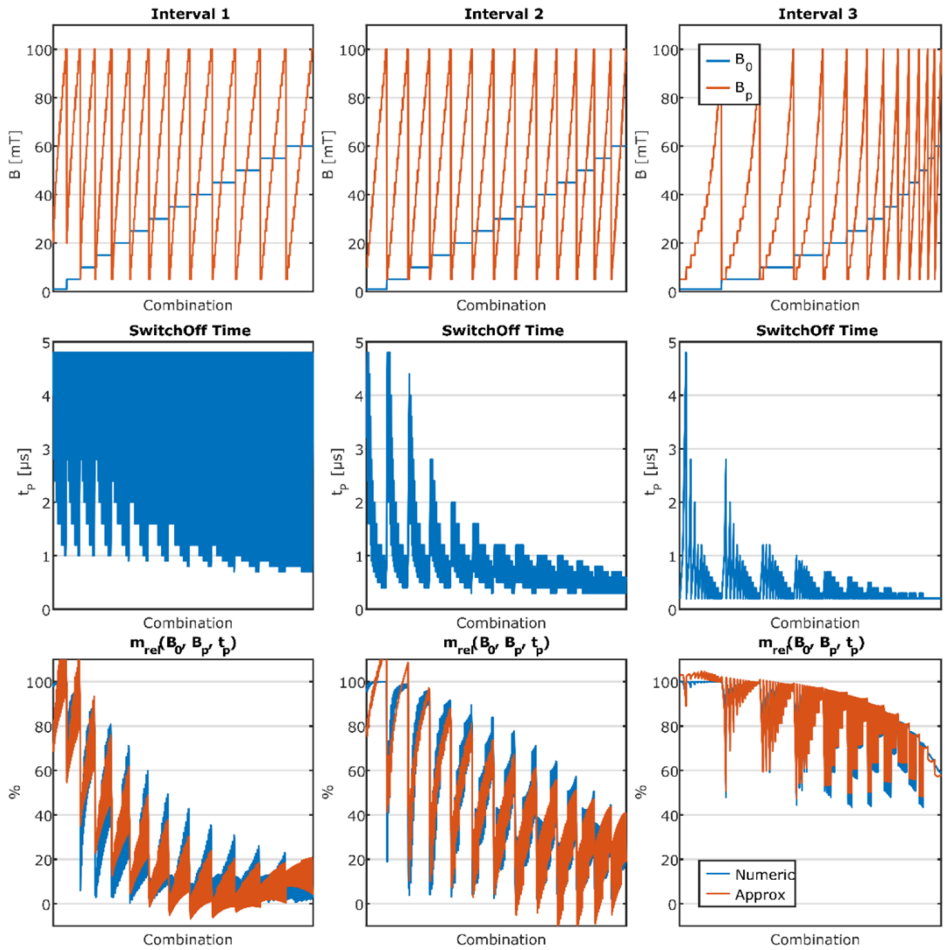

**Figure S1:** Comparison of the approximated and numerically calculated solutions for $m_{rel}$ depending on the various combinations of $B_0$, $B_p$, and $t_p$.

ii. Calculation of the signal after switch-off

Subsequent to the described switch-off process, the values of $B_x$ and $B_y$ are 0 and $B_z=B_0$. Consequently, the measurable transverse magnetization will rotate around the z-axis of the laboratory system. The Rx coil is oriented along the y-axis. To calculate the induced voltage the reciprocity principle can be used [3]. For the purposes of simplicity, the following relationship is assumed for the transverse magnetization $M_\perp$ for $t > t_p$ (with parameterization $t' = t - t_p$ and a start phase $\vartheta$):

$$M_\perp(t') = M_\perp(t_p) exp\left(t'/T_2^*\right) sin(\omega_L(B_0)t' + \vartheta) \qquad (S2.1)$$

where transverse relaxation was replaced by $T_2^*$ due to the expected inhomogeneities coming from the switching field and its disturbances. This results in the induced voltage in a receiving coil being [3], [4]:

$$U(t') = \int_V \omega_L B_\perp(dV) M_\perp(t') \, dV \qquad (S2.2)$$

where the spatial sensitivity of the Rx coil is considered in the parameter $B_\perp(dV)$.

In the ideal experiment, $B_x(t=0) \gg B_z$ and $t_p \to 0$ applies (extremely fast and strong switchoff) so that $m_{rel}$ becomes $m_{rel} \approx 1$. Then the remaining transverse magnetization after the switchoff is:

$$M_\perp(t_p) \approx M_{\perp 0}(B_0, B_p) = \rho_0 \frac{\gamma^2 \hbar^2}{4kT} B_p \qquad (S2.3)$$

And thus, the induced voltage becomes (without relaxation processes):

$$U(t') = \gamma B_0 M_\perp(t_p) exp\left(t'/T_2^*\right) sin(\omega_L(B_0)t' + \vartheta) \int_V B_\perp(dV) \, dV \sim B_0 B_p \qquad (S2.4)$$

iii. Simulated Noise for SNR Estimation

As the only noise source, the thermal noise of the ohmic resistance *R* of the Rx-coil was used in simulation. Therefore, the frequency dependent Skin- and Proximity effect were considered, which significantly increase the ohmic DC-resistance $R_{DC}$ depending on the chosen coil geometry and measured frequency of the signal *f*.

For the skin-effect a frequency dependent factor $\Delta_{sk}(f) > 1$ [5] and for the Proximity-effect a second frequency dependent factor $\Delta_{px}(f) > 1$ [6] is calculated, which results in an effective ohmic resistance of $R(f) = \Delta_{sk}(f)\Delta_{px}(f)R_{DC}$.

The thermal noise $U_{noise}$ of this resistance is calculated using the Johnson-Nyquist formula:

$$U_{noise} = \sqrt{4\,k_B\,T\,R(f)\,BW} \qquad (S3.1)$$

with the Boltzmann constant $k_B$, the temperature *T*=300K, the ohmic resistance *R(f)* of the Rx-coil and the bandwidth *BW* of the Rx electronic.

For $B_0 \leq 5mT$ the sampling frequency was set to $f_S = 2MSps$ and therefore the sampling bandwidth is $BW = f_s/2 = 1MHz$, resp. for $B_0 = 45mT$ the sampling frequency was set to $f_S = 10MSps$ and therefore the sampling bandwidth is $BW = f_s/2 = 5MHz$. The cut-off-frequency of the anti-aliasing-filter was held constant at $f_{-3dB} = 3MHz$, which therefore results in aliasing for $B_0 \leq 5mT$ and increased effective bandwidth to $BW = 1.95MHz$ and therefore increased noise. For $B_0 = 45mT$ the anti-aliasing-filter limits the effective bandwidth to $BW \approx 1.006 \cdot 3MHz$ (since a 6[th] order Butterworth filter is used).

The SNR was then calculated using the effective value of the simulated signal $u_{sig} = \hat{u}_{sig}/\sqrt{2}$:

$$SNR = 20\,log_{10}\,u_{sig}/U_{noise} \qquad (S3.2)$$

iv. Long-term investigation of Aging of the Silicon Carbide Field-Effect Transistors (SiC-nFET)

In the presented operating mode of the SiC-nFETs, rapid switch-off of the polarization coil is enabled via a triggered avalanche of the source-drain barrier of the FET. This involves applying high power to the SiC crystal for short periods of time, which can cause long-term damage to the FET.

The ageing of the SiC-nFETs was characterized using a separate setup. The adaption of the polarization circuit involved the integration of a single SiC-nFET, with the configuration set to $R_S = 1k\Omega$ and $R_D = 225\Omega$, thereby ensuring the reliable avalanche-event (modified switchnode capacitances due to lower number of installed SiC-nFETs). The original polarization coil ($L = 80\mu H$) was taken as the inductive load. The SiC-nFET is electrically connected with a screw connection. The SiC-nFET was cooled via an aluminum heat sink with active air flow (cooler temperature in steady state: approx. $T = 35°C$). The coil was passively cooled in a water basin containing approx. $V = 5l$ of water at room temperature. A microcontroller-based sequencer (ATmega 328, Arduino Nano) switches the polarization generator on for $t = 1s$ and then off for $t = 10s$ (avalanche triggering and cooling phase). The current was set to $I = 13.44A$.

Following a predetermined number of breakdowns, the SiC-nFET was removed from the polarization circuit. The on-resistance $R_{DSon}$, threshold voltage $V_{th}$, reverse current $I_{DSS}$, and capacitances $C_{iss}$, $C_{oss}$ and $C_{rss}$ were then determined using a curve tracer (Keysight B1505A). The steps selected for analysis were 0, 100, 350, 750, 1,500, 2,500, 3,500, and 10,000. The results are shown in Figure S2. With the exception of a minor discrepancy due to measurement tolerance, no signs of aging or damage can be observed. In the standard measurement sequence employed for the field mapping (6s pre-polarization, 1s sampling, and neglecting the cooling of the sample), the result of 10,000 avalanches would be a continuous measurement time of almost 20 hours, or 60 hours when considering the three installed transistors. In principle, after this period, the SiC-nFETs could be replaced as a preventative measure, or an additional measuring circuit could be employed to measure the reverse current and indicate any degradation.

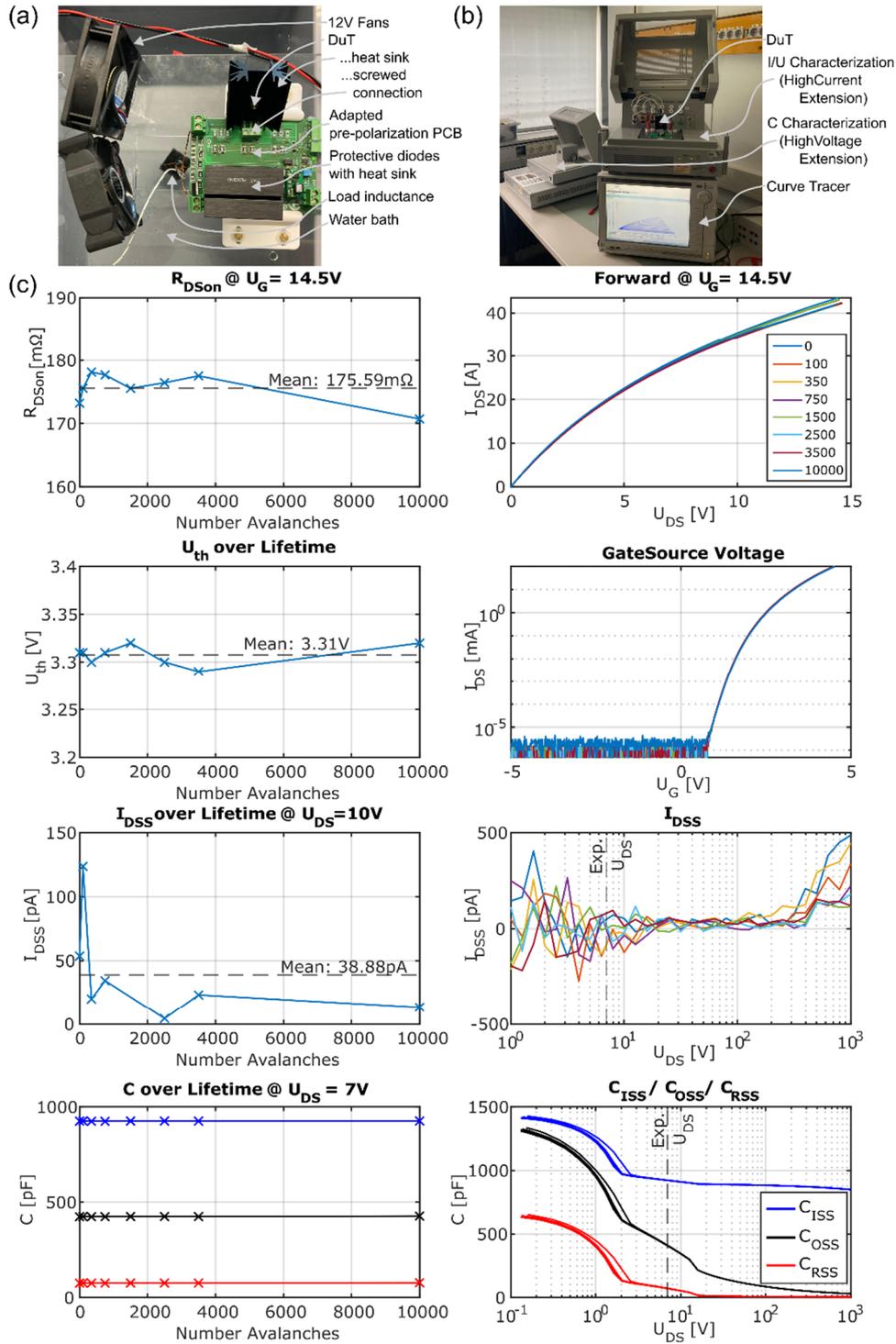

**Figure S2:** Aging test of the SiC nFET used. **(a)** Test setup during consecutive breakdowns (analogous to the sensor concept). **(b)** Measurement of the dissembled SiC-nFET with a curve tracer. **(c)** Measurement results with the measured traces (right side) and the derived parameters (left side).

## v. Calculation of the normalized slope (Figure 2e)

We compared the evolution of the maximum amplitude in the NMR spectrum $\hat{u}$, or the *SNR*, across varying $B_p$ field strengths ($t_p$=6s fixed). The measured $\hat{u}(B_p)$, resp. *SNR*($B_p$), is shown in Figure 6a. The simulated values are shown in Figure 2d. Due to slight variations in the Rx coil geometries, a direct qualitative comparison between the two measurements is not feasible (the signal amplitude is significantly smaller in the measurement due to less windings).

Because of the random winding pattern of the real Rx coil, modeling it accurately in simulation is not possible.

To ensure a quantitative comparison, the values were normalized to the maximum value before a linear fit (both for the measured and simulated data):

$$\hat{u}_{norm}(B_p) = \hat{u}(B_p) / \max_{\forall B_p} \hat{u}(B_p) \tag{S5.1}$$

$$SNR_{norm}(B_p) = SNR(B_p) / \max_{\forall B_p} SNR(B_p) \tag{S5.2}$$

Note that *SNR* is used here in its absolute, linear form (**not** logarithmic in dB). These normalized data are then linearly fitted (least-square fit method), resulting in the following relationship:

$$\tilde{u}(B_p) = a_{\hat{u}} B_p + OS_{\hat{u}} \tag{S5.3}$$

$$\widetilde{SNR}(B_p) = a_{SNR} B_p + OS_{SNR} \tag{S5.4}$$

with the slopes $a_{\hat{u}}$, resp. $a_{SNR}$, (unit 1/mT) and a dimensionless offset $OS_{\hat{u}}$, resp. $OS_{SNR}$. Both slopes are shown in Figure 2e of the paper.

The tolerance of the fitted slope is calculated from the calculated residuals. For the simulated SNR this becomes very small with error bars almost not visible in the scaling of Figure 2e.

vi.  Analysis Gradient Measurements (Figure 7)

In the experiments with the switched-on DC-gradients in Figure 7, the frontend remains stationary: All three measurement series per $B_0$ field strength (each with either the X-, Y-, or Z-gradient active) are made at the same location (the frontend was not removed or moved in between). As the position was not moved by known offsets (e.g., using a positioning robot), the exact spatially resolved efficiency of the gradients cannot be determined. However, this is not necessary for use in a field camera for improved reconstruction (e.g. as in [7]). However, when using the sensor as a field camera, it is crucial to know the exact position of the sensor relative to the MRI scanner, resp. the gradient centers.

For this measurement, the sensor was positioned as close as possible to the isocenter of the $B_0$-magnet (for maximum homogeneity). The gradients were operated with an adjustable DC current $I_G$ ($-250mA \leq I_G \leq 250mA$). Each gradient was activated individually (only one switched-on at a time). The simulated efficiency of the gradient system ($\tilde{G}_X = \tilde{G}_Y = 1594 nT/m/A$, $\tilde{G}_Z = 4833 nT/m/A$) can be used to calculate the applied gradient $G_{X,Y,Z} = \tilde{G}_{X,Y,Z} I_G$ (shown as the x-axis of the plot) by multiplying it by the measured current $I_G$.

The frequency drift $\Delta f(G_{X,Y,Z})$ relative to the zero point $f_0 = f_{X,Y,Z}(G_{X,Y,Z} = 0 nT/cm) = \gamma B_0$ (all gradients switched off) was calculated from the acquired NMR spectra:

$$\Delta f_{X,Y,Z}(G_{X,Y,Z}) = f_{X,Y,Z}(G_{X,Y,Z}) - f_0 \qquad (S6.1)$$

This frequency drift was linearly fitted:

$$\widetilde{\Delta f}_{X,Y,Z}(G_{X,Y,Z}) = a G_{X,Y,Z} + OS \qquad (S6.2)$$

with the slope $a$ and an offset $OS$.

This linear fit can be used to determine the distance $d_{X,Y,Z}$ to the isocenter of the individual gradient:

$$d_{X,Y,Z} = \frac{\widetilde{\Delta f}_{X,Y,Z}(G_{X,Y,Z})}{\gamma G_{X,Y,Z}} \qquad (S6.3)$$

The results are shown in Figure 7. Note that the isocenters of the three individual gradients in the ULF setup do not lie perfectly on top of each other, as they have not been perfectly aligned mechanically.

vii.   Investigation of electronic induced time delays after completion of pre-polarization

In consideration of the extremely short FID in the Halbach magnet ($T_2^* \ll 1ms$), it is essential for maximum sensitivity that sampling begins immediately after the non-adiabatic switch-off has ended. However, the receive path must be enabled as latest as possible in order to protect it from strong induced transients due to the switch-off. In this setup, it is necessary to consider the reasons for possible delays on the receive side and the pre-polarization side separately in order to minimize them.

Receive Electronics

The input of the first LNA is short-circuited via an photocoupler (TLP3431, Toshiba Electronic, Tokyo, Japan) during the pre-polarization phase to protect the receive path. The active HIGH state indicates that the photocoupler is conducting. Therefore, the trigger must fall so that the LNA can receive a signal. The remaining electronics (VGA, filter, ADC) are activated at all times. The photocoupler control line is filtered via an RC low-pass filter and a voltage shifter, positioned directly after the sequencer and directly before the photocoupler. Additionally, the photocoupler has a specified design-related turn-off time.

This total delay was measured experimentally by feeding a sine wave signal ($f = 1.7MHz$) of a frequency generator via a coaxial connection into the frontend input instead of the Rx coil. The RF signal was captured directly after the photocoupler using an installed coaxial connector and an oscilloscope. Furthermore, the trigger was recorded with the oscilloscope connected directly after the sequencer. This results in a total delay of $t_d \approx 1.1ms$ between the falling trigger edge and beginning of the signal measurement (see Figure S3a). This delay is consistent across various experiments and can be accommodated by adjusting the sequence (trigger is switched off slightly earlier than it should).

Pre-Polarization Circuit

The polarization circuit is controlled by a trigger from the sequencer. Active HIGH indicates that the pre-polarization is active (DC current $I_p$ is flowing). When the trigger is deactivated, the non-adiabatic switch-off process is initiated, with a slope according to the installed setup on the PCB. The trigger is filtered using a second-order RC low-pass filter and a voltage shifter. The ACPL-P343 (Broadcom Inc, San José, USA) gate driver ensures galvanic isolation via an integrated optocoupler.

All three SiC nFETs are controlled via a single $R_G = 1\Omega$ each. The gate driver provides a $U_G = +15V$ when it is switched on and a $U_G = 0V$ when it is switched off. Together with the gate capacitance of the SiC-nFETs, this results in a total delay between the falling trigger edge and the transistor actually being switched off. To measure the delay, the signal was captured directly at the gate/source pads of a SiC-nFET using a twisted wire and fed out to a BNC connector (with serial resistors $R = 330k\Omega$ (Source) und $R = 1M\Omega$ (Gate) to minimize capacitive loading). The galvanic isolation process was carried out using a low-budget differential probe (model DP750-100, Micsig, Shenzen, China). The measurement results in a total delay of $t_d \approx 8.8ms$ (see Figure S3b).

Adapted Sequence for Low Field Measurement at $B_0$ = 45mT

It is evident that these delays pose significant challenges to automated measurement of FIDs, particularly in (inhomogeneous) Halbach magnets. Therefore, the sequence was adjusted, resp. calibrated for the possible delays, so that the triggers are switched off before their actual determined time point to overcome the measured delays. Subsequently, in 20 consecutive experiments, the resulting delay between the switch-off off of the pre-polarization circuit and the switch off of the photocoupler (LNA starts amplification) was measured (see Figure S3c). Further tolerances were identified, with an overall delay ranging from $2\ \mu s < t_d < 26 \mu s$ (see Figure S3d).

In order to characterize the sensor in low field with short $T_2^*$, the gate/source voltage was also sampled. The oscilloscope's data acquisition was initiated by the falling edge of the receive trigger. In post-processing, the measured delay of the falling edge of the gate/source voltage was then evaluated. An offset of $\Delta = 1\mu s$ was then added to this time point (where the avalanche-induced collapse of $B_p$ occurs), and data analysis was initiated from subsequent measurement points onwards.

Sequence for Ultra Low Field Measurement at B$_0$ = 1 to 5 mT

Due to the long FID with $T_2^* > 50 ms$, the original sequence was used without further adjustments or consideration of existing delays. During the course of all measurements at all fields, no damage to the LNA was detected.

In post-processing, only data for $t \geq 7.5 ms$ resulting from the measured delays ($t_d \approx 8.8ms - 1.1ms$) is evaluated. The additional apodization filter ensures that fluctuations in measurement points at the start of evaluation have minimal impact. In a subsequent development, however, a hardware delay should also be installed for this operating case at ULF.

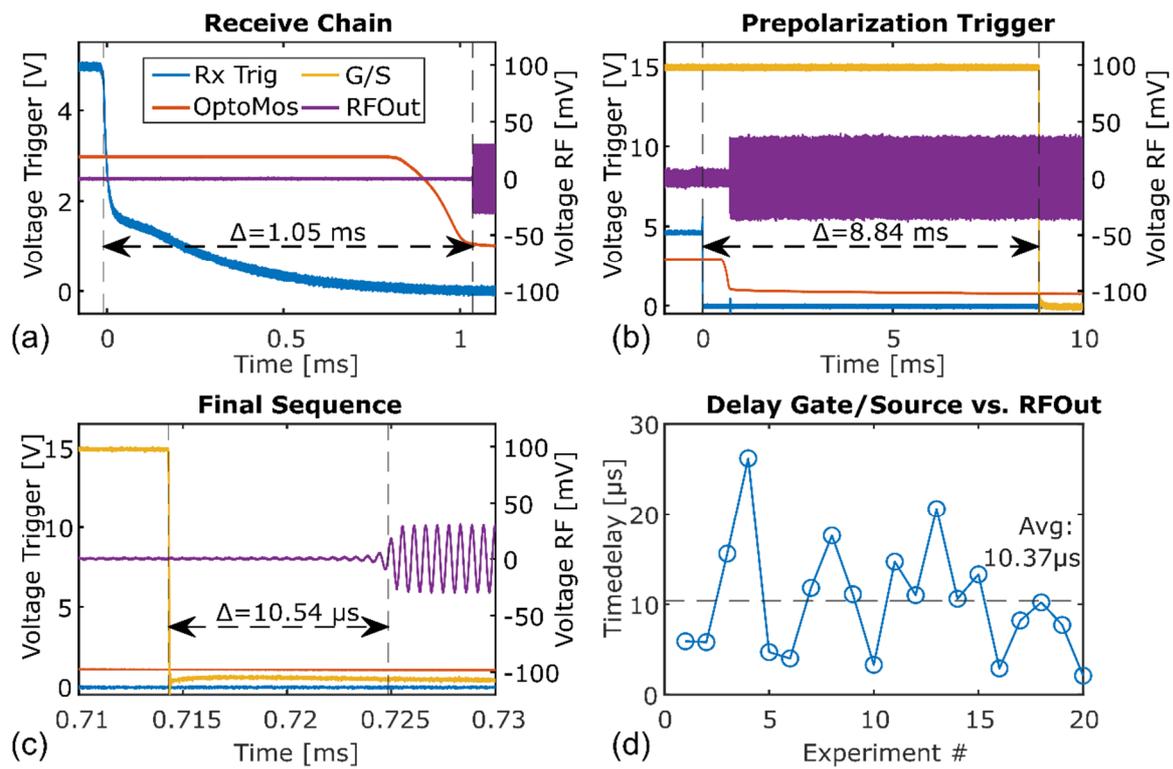

**Figure S3:** Measured delays in the assembled setup. "Rx Trig" stands for the trigger output of the sequencer that controls the photocoupler (On=Input LNA short-circuited). "OptoMos" is the control input directly at the photocoupler (i.e., the filtered "Rx Trig" signal). "G/S" is the gate/source voltage directly at the SiC-nFET pins. "RF Out" is the RF signal line between the photocoupler and the LNA. If the voltage here is ~0mV, the LNA is effectively protected. **(a)** Investigation of receive electronics. **(b)** Investigation of the pre-polarization circuit. **(c)** Investigation of the final sequence for the LF setup (Halbach measurement). **(d)** Measured variation in delay between different experiments for the LF setup.

## viii. Investigation of Power Source Instability for the Ultra Low Field magnet

In addition to the inhomogeneity of the magnetic field, the current drift caused by suboptimal control of the current source has a significant impact on the measurement results of the sensor presented. In order to verify the current drift over a longer period of time, the used BCS5/75 current source (75V/5A, HighFinesse GmbH, Tübingen, Germany) was connected to a Helmholtz-like configuration with four coils, and the NMR peak of a water sample was measured with a SQUID sensor [8]. The field was set first to $B_0 = 0.5 mT$ and then to $B_0 = 1 mT$. The low-pass behaviour of the SQUID electronics and the cooling capacity of the $B_0$ coil preclude the possibility of higher fields. The SQUID, being operated within a cryostat at a temperature of $4.2 K$, is not directly applicable to the build ULF solenoid coil (see SI chapter 10). Consequently, the determined current drift is scaled to the solenoid coil used and should therefore be understood as an indication of the actual current drift (assuming that the different inductance has a negligible effect on the control of the current source). Each measurement point was recorded at a rate of $t_R = 12s$, with a total of 50 sequential measurements being acquired. Two measurement series were recorded at $B_0 = 0.5 mT$ ($I = 0.762 A$) and $B_0 = 1 mT$ ($I = 1.506 A$). This results in a field drift of $\Delta f = -13.173 mHz/A/s$ for the measurement (which means a controller inaccuracy of $\Delta I = -0.47 ppm/s$) (see Figure S4). When scaled to the solenoid coil, this results in $\Delta f = -42.775\ mHz/A/s$. For a standard sensor measurement of $t = 10s$ (2s Hall sensor, 6s prepolarization, sampling, and pause), the resultant drift per measurement is $0.2 Hz$ (1mT), $0.6 Hz$ (3mT) and $1 Hz$ (5mT) respectively.

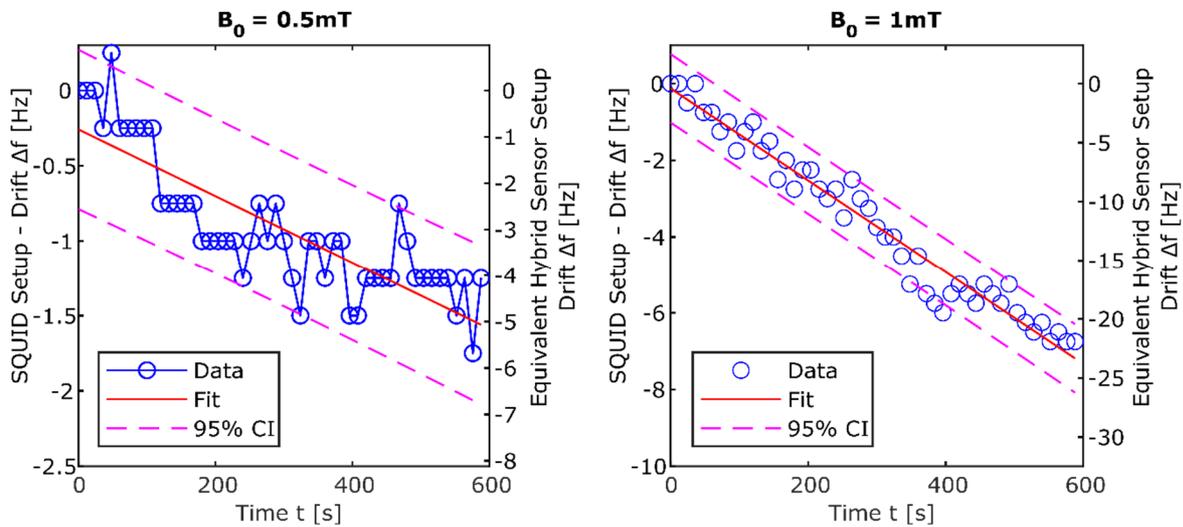

**Figure S4:** Drift of the current source measured with the SQUID setup for the field $B_0 = 0.5 mT$ and $B_0 = 1 mT$.

ix. Accuracy of frequency measurement with the NMR field probe

To determine the accuracy of the sensor, we use the bin width of the discrete Fast Fourier Transform (FFT). Ideally, the measured NMR signal is an oscillating signal *u(t)* with Larmor frequency $f_0$, that is amplitude-modulated by an exponentially decaying function with time constant $\tau$:

$$u(t) = \hat{u}\, sin(2\pi f_0 t)\, exp(-t/\tau) \qquad (S9.1)$$

To estimate the accuracy of the sensor, we assume that the signal cannot be detected after a time $t_0 = 3\tau$ ($u(t_0)/\hat{u} \approx 5\%$, signals with a $SNR > 27dB$ should be significantly above the noise floor and detectable therefore). The following upper limit is calculated for the bin width of the FFT:

$$Bin = \frac{1}{3\tau} \qquad (S9.2)$$

and thus, the maximum resolution of the frequency analysis is:

$$\Delta f = \frac{1 Bin}{f_0} \qquad (S9.3)$$

If the *SNR* is significantly worse (as in the case of the measurement at $B_0 = 45mT$), $t_0$ should be reduced to $t_0 = 2\tau$ ($SNR > 18dB$).

For the evaluation of the measurements, $\tau = T_2^*$ was set and the measured relaxation times $T_2^*$ were used. This results in an estimated maximum accuracy of $\Delta f$:

**Table S1:** Calculated accuracy of the sensor for the different magnetic fields and the measured relaxation times.

| Measurement | $B_0$ | Δf |
| --- | --- | --- |
| Figure 5 | 1mT | 235ppm |
| Figure 5 | 3mT | 26ppm |
| Figure 5 | 5mT | 16ppm |
| Figure 9 | 45mT | 1051ppm (0.1%) |

x. Ultra Low Field Magnet

It was determined that the approach used to develop the Ultra-Low Field MRI (ULF MRI) for this work, facilitates the construction of an easy to replicate magnet and gradient system for ULF MRI, which is both cost-effective and functional (see detailed images in Figure S5). This system has the potential to be particularly advantageous in the context of workshops and for the training of students. It is therefore the intention of this chapter to provide more detailed information on the subject, with reference to the CAD files, which are available in the GitHub repository.

The central $B_0$ magnet was optimized analytically using Biot-Savart's law (see scripts provided in [9]) and can be calculated on a standard office PC in a few seconds. In the case of the solenoid diameter being substantially larger than the volume to be examined, and the solenoid length being significantly longer than the diameter, a highly homogeneous magnetic field can be assumed (in the present case, $12\ ppm$ for a spherical FOV with a diameter of $20\ mm$). Manufacturing tolerances resulting from the manual winding of the magnet are largely negligible. It is to be expected that minor variations in the number of wound turns will occur due to the tolerance of the copper wire diameter. It is recommended that the optimization of the second layer should be done only subsequent to the completion of the first layer. Additional multi-stage shimming steps, as is the case with Halbach-based systems, are not needed. Any type of cylinder can be utilized as a winding body. In this instance, a Plexiglas cylinder was utilized. However, good results have also been achieved with off-the-shelf drain pipes. Due to electrical power loss, the field is usually limited to $B < 10mT$, although the polarization can be significantly increased by using (not particularly homogeneous) pre-polarization coils (e.g., 70mT is possible in our sensor).

The gradients are straightforward to produce due to their planar structure.

The gradients in the X and Y directions (which exhibit an identical structure, but are rotated by 90° relative to each other) are readily producible. Essentially, these gradients comprise a set of four coils, with a rectangular base area in each spatial direction. The eight corner points of a coil were calculated using a Biot-Savart-based optimization so that the gradient in the FOV (in this case, a spherical FOV with a diameter of $100mm$) is as linear as possible. These eight points were then transferred to a Plexiglas disc ($10mm$ thick). An M10 thread was cut at this point, and a nylon screw was screwed in, leaving approx. $5mm$ of the thread exposed. The enameled copper wire was then simply wound around these corner points. The z-gradient has been designed as a Maxwell coil, which, in principle, geometrically corresponds to a Helmholtz coil with a slightly increased distance between the coils of $d = \sqrt{3}R$ (with the goal of as linear a gradient as possible). In contradistinction to the Helmholtz coil, the two coils are wound in opposite directions with copper wire. In this instance, the winding body comprised Plexiglas rings, with two covers affixed by screws to a central core. The whole gradient system has been adapted from a ready-to-use ULF MRI setup [8]. In order to ensure similar results to the simulation, the individual coils of the gradients must be connected in a way that there is only one connection wire for the power sources. When laying the wires, care must be taken to ensure that two wires with opposite currents (forward and return) are always run in parallel so that the magnetic fields of the current conductors cancel each other out macroscopically.

Established open-source power sources ($I \leq 8A$) can be utilized as power sources for the gradients:

opensourceimaging.org/project/current-driver-for-local-b0-shim-coils/.

The power source for the $B_0$ coil must have as little drift as possible, whereby the large inductive load contributes to the stability (series resistors may be necessary to stabilize the current regulator of the source). The open-source power source, which has been published for some time now, ($I \leq 5A$):

opensourceimaging.org/project/eez-h24005-bench-power-supply/

is suitable here.

If a pre-polarization coil is installed, it can be operated with any available power supply (no requirements for stability and accuracy).

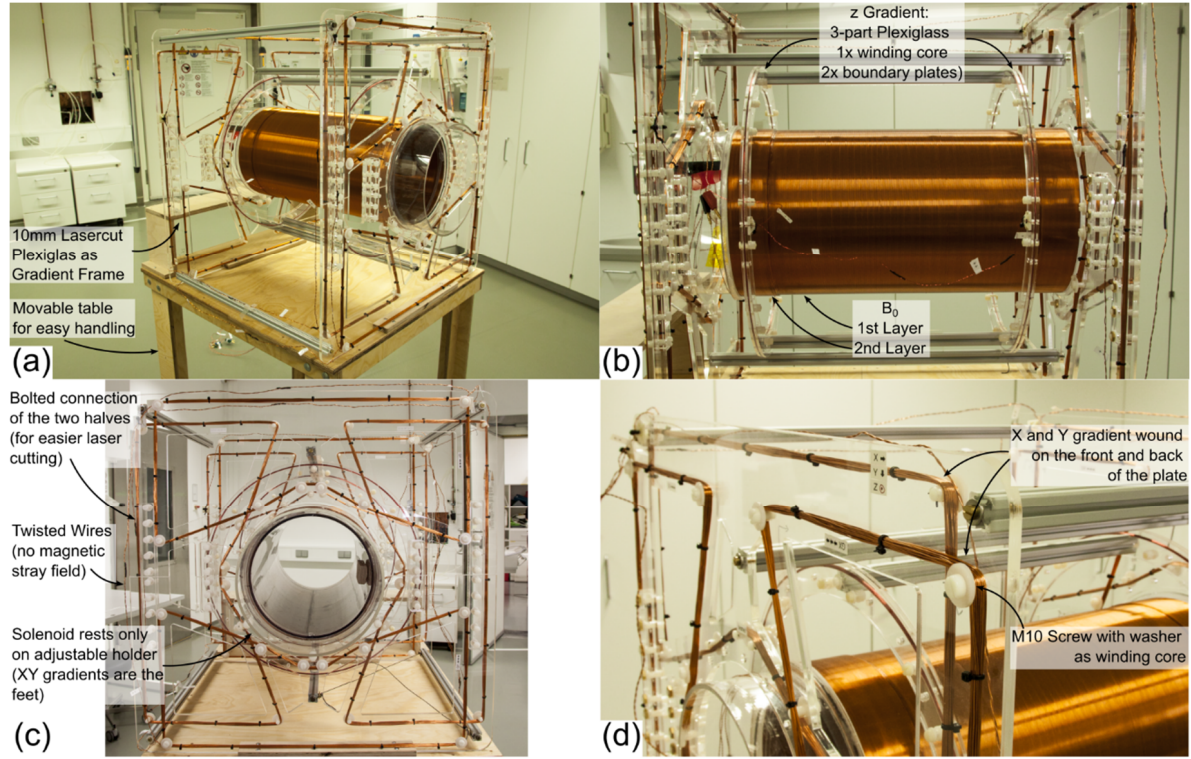

**Figure S5:** Detailed views of the assembled ULF setup. (a) Perspective overview with easily movable table. (b) Side view of B0 magnet with 2 coil layers. (c) Front view of magnet. (d) Detailed view of XY gradient winding.

## xi. Supplementary Information References